\documentclass[onefignum,onetabnum]{siamart171218}

\usepackage{lipsum}
\usepackage{amsfonts}
\usepackage{graphicx}
\usepackage{epstopdf}

\ifpdf
  \DeclareGraphicsExtensions{.eps,.pdf,.png,.jpg}
\else
  \DeclareGraphicsExtensions{.eps}
\fi
\usepackage{amssymb}
\usepackage{amsmath}
\usepackage{xcolor}
\usepackage{enumitem}
\usepackage{listings}
\usepackage{url}
\usepackage{multirow}

\lstdefinestyle{interfaces}{
	float,
	floatplacement=tbp
}
\usepackage{textcomp}
\usepackage{fancyhdr}
\usepackage[author={Mustafa}]{pdfcomment}
\usepackage[utf8]{inputenc}
\usepackage[english]{babel}
\usepackage{mathtools}
\usepackage{setspace}

\usepackage{subcaption}
\usepackage{cleveref}

\graphicspath{{figs/}}

\newsiamremark{remark}{Remark}
\newsiamremark{hypothesis}{Hypothesis}
\crefname{hypothesis}{Hypothesis}{Hypotheses}
\newsiamthm{claim}{Claim}

\headers{
  Extreme Scale FMM-Accelerated BIE Solver
}{M. Abduljabbar, M. Al Farhan, \emph{et~al.}}

\title{
  Extreme Scale FMM-Accelerated Boundary Integral Equation Solver for Wave Scattering
}

\author {
  Mustafa Abduljabbar\thanks{King Abdullah University of Science and Technology, Thuwal, Saudi Arabia.}
  \and Mohammed Al Farhan\footnotemark[1]
  \and Noha Al-Harthi\footnotemark[1]
  \and Rui Chen\footnotemark[1]
  \and Rio Yokota\thanks{Tokyo Institute of Technology, Tokyo, Japan.}
  \and Hakan BAGCI\footnotemark[1]
  \and David Keyes\footnotemark[1]
}

\usepackage{amsopn}

\begin{document}

\maketitle

\begin{abstract}
Algorithmic and architecture-oriented optimizations are essential for achieving performance worthy of anticipated energy-austere exascale systems. In this paper, we present an extreme scale FMM-accelerated boundary integral equation solver for wave scattering, which uses FMM as a matrix-vector multiplication inside the GMRES iterative method. Our FMM Helmholtz kernels are capable of treating nontrivial singular and near-field integration points. We implement highly optimized kernels for both shared and distributed memory, targeting emerging Intel extreme performance HPC architectures. We extract the potential thread- and data-level parallelism of the key Helmholtz kernels of FMM. Our application code is well optimized to exploit the AVX-512 SIMD units of Intel Skylake and Knights Landing architectures. We provide different performance models for tuning the task-based tree traversal implementation of FMM, and develop optimal architecture-specific and algorithm aware partitioning, load balancing, and communication reducing mechanisms to scale up to 6,144 compute nodes of a Cray XC40 with 196,608 hardware cores. With shared memory optimizations, we achieve roughly 77\% of peak single precision floating point performance of a 56-core Skylake processor, and on average 60\% of peak single precision floating point performance of a 72-core KNL. These numbers represent nearly 5.4x and 10x speedup on Skylake and KNL, respectively, compared to the the baseline scalar code. With distributed memory optimizations, on the other hand, we report near-optimal efficiency in the weak scalability study with respect to both the $O(\log{P})$ communication complexity as well as the theoretical scaling complexity of FMM. In addition, we exhibit up to 85\% efficiency in strong scaling. We compute in excess of 2 billion DoF on the full-scale of the Cray XC40 supercomputer. The numerical results match the analytical solution with convergence at 1.0e-4 relative 2-norm residual accuracy. To the best of our knowledge, this work presents the fastest and the most scalable FMM-accelerated linear solver for oscillatory kernels.

\end{abstract}

\begin{keyword}
3D Helmholtz equation,
%Acoustics wave scattering,
%Communication reduction,
Data-level parallelism,
%Domain-decomposition,
%Extreme scale,
Fast Multipole Method,
$\mathcal{HSDX}$,
Load balancing,
Thread-level parallelism
\end{keyword}

\begin{AMS}
%%%%%%%%%%%%%%%%%%%%%%%%%%%%%%%%%%%%%%%%%%%%%%%%%%%%%%%
%%%%%%%%%%%%%%%%%%%%%%%%%%%%%%%%%%%%%%%%%%%%%%%%%%%%%%%
%%%%%%%%%%%%%%%%%%%%%%%%%%%%%%%%%%%%%%%%%%%%%%%%%%%%%%%
%%%%%%%%%%%%%%%%%%%%%%%%%%%%%%%%%%%%%%%%%%%%%%%%%%%%%%%
\end{AMS}

%%%%%%%%%%%%%%%%%%%%%%%%%%%%%%%%%%%%%%%%%%%%%%%%%%%%%%%%%%%%%%%%%%%%%%%%%%%%%%%%%%%%%
%%%%%%%%%%%%%%%%%%%%%%%%%%%%%%%%%%%%%%%%%%%%%%%%%%%%%%%%%%%%%%%%%%%%%%%%%%%%%%%%%%%%%

\section{Introduction}
\label{sec:introduction}

%Waves are oscillatory phenomena that transfer energy though a medium, vacuum or non-vacuum.
%Such energy transfer might deviate from its incident direction due to localized nonuniformity in the medium \cite{NohaRef1Sheng95},
%which is so-called a ``scattering object.'' Acoustic wave scattering, which is driven by arbitrary scatterers that are embedded in a 3D medium,
%is one of the key consequences of the nonuniformity of the propagation medium \cite{NohaRef2Bohren2007}.
%It is a very important phenomena of interest in many computational science \& engineering (CS\&E) applications,
%such as underwater sonar acoustics, aeronautics \cite{Rudi:2015:EIS:2807591.2807675}, high-intensity focused ultrasound \cite{Betcke2017}, etc.
The Boundary Element Method (BEM) \cite{NohaRef3Williams94} is one of the most efficient means to compute the numerical solution of the wave scattering problem through relying upon dimensionality reduction.
Following the surface discretization, the resulting linear system of equations ($Ax=b$) features a dense matrix.
In order to efficiently solve such a linear system, one can use an iterative method, e.g., GMRES,
%\cite{doi:10.1137/0907058}
inside which
a fast hierarchical approximation of the matrix-vector product can be computed through utilizing Fast Multipole Method (FMM) \cite{Gumerov2009}.
Since FMM uses for nonuniform grids an octree-based hierarchical domain decomposition,
%with different computational structures similar to the Geometric Multigrid (GMG),
it exhibits near-optimal scalability \cite{Gholami2016}.
In addition, the Multilevel Fast Multipole Algorithm (MFLMA) employs FMM as a low frequency accelerator for the matrix-vector multiplication kernel.
Essentially, MFLMA can be used as an efficient solver to compute integral equations of electromagnetic \cite{Michiels2015} or acoustic wave scattering \cite{Hao2015}.

%Boundary Element Method (BEM) \cite{NohaRef3Williams94} is one of the most efficient means to compute the numerical solution of the wave scattering problem through relying upon dimensionality reduction.
%Following the surface discretization, the resulting linear system of equations ($Ax=b$) features a dense matrix.
%In order to efficiently solve such a linear system, one can use an iterative method, e.g., GMRES \cite{doi:10.1137/0907058}, inside which
%a fast hierarchical approximation of the matrix-vector product can be computed through utilizing either: 1) the Fast Multipole Method (FMM) \cite{Gumerov2009},
%or 2) its algebraic variant ($\mathcal{H}$-matrices).
%Since FMM uses for nonuniform grids an octree-based hierarchical domain decomposition with different computational structures similar to the Geometric Multigrid %(GMG), it exhibits near-optimal scalability \cite{Gholami2016}.
%Essentially, FMM can be employed either as a: 1) linear preconditioner \cite{Takahashi2017}, or
%2) low frequency accelerator for the matrix-vector multiplication kernel (Multilevel Fast Multipole Algorithm (MFLMA)).
%In particular, the MFLMA is known for solving integral equations of electromagnetic
%\cite{Michiels2015} or acoustic wave scattering \cite{Hao2015}.
 
In this paper, we present an extreme scale, rapidly converging implementation of 
an FMM-accelerated linear solver for wave scattering for the complex 3D Helmholtz Boundary Integral Equation (BIE).
FMM is a very compute intensive algorithm \cite{FMM:KNC} that is portable and adaptable
to different levels of parallelism \cite{Abduljabbar2017_1}, and exhibits a scalable communication \cite{Abduljabbar2017_2,zandifar2015}.
It is thus natural to rely upon such algorithm to accelerate the matrix-vector multiplication kernel to scale
the application performance to a large number of tightly-coupled compute nodes.
To the best of our knowledge, the underlying 3D Helmholtz kernel that is described in \cite{Malhotra2016} does not possess a self-contained near or self-singularity treatments, through which the accuracy of the solution irrespective of the convergence rate is improved \cite{NohaRef20:doi:10.1137/0719090}.
Our solver, however, handles both self-contained near and self-singularity treatments, while efficiently contending with 
 a wide range of extreme scale performance challenges.
A direct application of our solver, which is experimented herein, is the acoustic wave scattering that
is driven by arbitrary scattering objects embedded in a 3D medium \cite{NohaRef1Sheng95, NohaRef2Bohren2007}. 
It is a very important phenomena of interest in many computational science \& engineering (CS\&E) applications,
such as underwater sonar acoustics, aeronautics \cite{Rudi:2015:EIS:2807591.2807675}, high-intensity focused ultrasound \cite{Betcke2017}, etc.

The rest of the paper is organized as follows. In Section~\ref{sec:background}, we provide background on the underlying physics and mathematics of the application code. Section~\ref{sec:shared-memory}
presents the shared-memory optimization means implemented to speedup the single node performance.
In Section~\ref{sec:distributed-memory}, we describe the extreme scale
implementation aspects, which include partitioning, load balancing, and communication reducing.
Section~\ref{sec:Experiments} details for experimental reproducibility
the workload characterizations, underlying hardware and software
stack, and methodologies of designing the performance analysis.
Section~\ref{sec:results} presents our performance evaluation results.
% Then, in Section~\ref{sec:related_work}, we outline briefly the related work in terms
% of FMM, acoustic wave scatter, and 3D Helmholtz BIE.
Finally, we conclude in Section~\ref{sec:conclusion} with brief outline of our ongoing work.

%%%%%%%%%%%%%%%%%%%%%%%%%%%%%%%%%%%%%%%%%%%%%%%%%%%%%%%%%%%%%%%%%%%%%%%%%%%%%%%%%%%%%
%%%%%%%%%%%%%%%%%%%%%%%%%%%%%%%%%%%%%%%%%%%%%%%%%%%%%%%%%%%%%%%%%%%%%%%%%%%%%%%%%%%%%

\section{Background}
\label{sec:background}

This section outlines the mathematical and physical aspects
of our FMM-Helmholtz solver, including the formulation of the
incident wave scattering, the singularity extraction and treatment through the Duffy transformation, and
a concise description of the underlying FMM kernels.

\subsection{Boundary Integral Equation Formulation}
Let $S$ represent the surface of a closed scatterer residing in a homogeneous medium. In general, fields of propagation waves in such medium are governed by Equation~\ref{eq1} \cite{NohaRef4},
where $U(r,t)$ is the unknown acoustic pressure, $c$ is the speed of sound, and $\nabla^2$ is the Laplacian operator. 

\begin{equation}\label{eq1}
\nabla^2 U(r,t) - \frac{1}{c^2} \frac{\partial}{\partial t^2} U(r,t) = 0 \text{, }
\end{equation}

Given the time-harmonic wave (i.e., $e^{-jwt}$), $U(r,t)$ has the form of Equation~\ref{eq2}.

\begin{equation} \label{eq2}
U(r,t) = Re[U_0(r)e^{-jwt}]
\end{equation}

Inserting Equation~\ref{eq2} into Equation~\ref{eq1} obtains the Helmholtz Equation~\ref{eq3},
where $k$ is the wave number.

\begin{equation}\label{eq3}
\nabla^2 U_0(r) + k^2 U_0(r) = 0
\end{equation}

Equation~\ref{eq4} is the surface integral result of plugging the second form of the Green’s theorem \cite{NohaRef5:kreyszig_norminton_1994} into Equation~\ref{eq3}. $p(r')$ is the pressure field at the source point $r'$, $q(r') = \frac{\partial p(r')}{\partial n'}$ is the velocity, $p^{inc}(r)$ represents the incident plane wave at the observation point $r$, and $G(r,r')$ is the scalar Green’s function of Equation~\ref{eq6},
where $R=|\boldsymbol{R}|=|r-r'|$ is the distance between source and observation points.

\begin{equation}\label{eq4}
p^{inc} (r) + \int_{S} [\frac{\partial G(r,r')}{\partial n'}p(r') - G(r,r')q(r')]dS' = \frac{1}{2} p(r), r \in S
\end{equation}

\begin{equation}\label{eq6}
G(r,r')=\frac{e^{jkR}}{4 \pi R}
\end{equation}

Equation~\ref{eq7} considers the soft boundary condition \cite{NohaRef6:doi:10.1063/1.4914797} with $p=0$ in Equation~\ref{eq4}.

\begin{equation}\label{eq7}
	\int_{S} G(r,r')q(r')=p^{inc}(r)
\end{equation}

\subsection{Discretization via Nyström Method}

To discretize the scatterer's surface, we first divide it into curvilinear triangular patches for higher-order geometry modeling. Then, a high-order Nyström method \cite{NohaRef7:931148} is used to expand the unknown surface velocity. Each curvilinear patch has $N_i$ interpolation points defined on the patch. The unknown velocity is expanded as an interpolation based on its values at those points given by Equation~\ref{eq8},
where $\vartheta^{-1}(r')$ is the Jacobian at $r'$, $L_{(i,n)} (\zeta,\eta)$ is the Lagrange interpolater at $r'$ calculated in a right triangle system ${u,v}$ \cite{NohaRef7:931148}, and $\{I\}_{(i,n)}$ is the set of unknown expansion coefficients at the $i^{th}$ interpolation point on the $n^{th}$ patch.

\begin{equation}\label{eq8}
q(r') = \sum_{n=1}^{N_p} \sum_{i=1}^{N_i} \vartheta^{-1}(r')L_{(i,n)}(\zeta,\eta) \{I\}_{(i,n)}
\end{equation}

Substituting Equation~\ref{eq8} into Equation~\ref{eq7} and applying the point-matching testing \cite{NohaRef7:931148} at the interpolation points leads to a final discretized matrix Equation~\ref{eq9}, where $[V^{inc}]_{j,m}=p^{inc}(r_{(j,m)})$.

\begin{equation}\label{eq9}
	ZI = V^{inc}
\end{equation} 

The entries of $Z$ are given by Equation~\ref{eq10}. $r_{(j,m)}$ is the $j^{th}$ interpolation point on the $m^{th}$ patch: $j=1,...,N_p$ and $m=1,...,N_i$.

\begin{equation}\label{eq10}
	[Z]_{(j,m)(i,n)}=\int_{\Delta n} G(r_{(j,m)},r') \vartheta^{-1}(r')L_{(i,n)}(\zeta,\eta) dr'
\end{equation}

%\subsection{Singularity Extraction and Treatment}
Numerical solution of the Singular Integral Equations (SIE) with the Nyström method requires an evaluation of the singular integrals either numerically or analytically. Several singularity treatment techniques have been proposed (e.g., polar coordinate transformation \cite{NohaRef18:Schwab:1992:NCS:2715380.2715541}, singularity subtraction technique \cite{NohaRef19:1573738}, and Duffy transformation \cite{NohaRef20:doi:10.1137/0719090}).
In this work, we use the Duffy transformation, since it works well with the weak $1/R$
singularity from Equation~\ref{eq6} \cite{NohaRef21:Mousavi2009}.

\subsection{FMM as an Accelerator for the Boundary Integral Equation}
From an algebraic perspective, FMM works as a matrix-free accelerator for the matrix-vector multiplication of certain matrices arising from the elliptic PDEs \cite{greengard1987} that satisfy Green's formula  (see Equation~\ref{eq6}) \cite{Ibeid2016}. The solution of the linear system involving $G(r,r')$ can produce the target vector field (i.e., $q(r')$ of Equation~\ref{eq4}) iteratively.
As mentioned, the BEM discretization matrix is dense, and thus,
the BIE evolves into a summation (Equation~\ref{eq8}) that is very costly to calculate. Equation~\ref{eq11} is an explanatory example that uses FMM to approximate the scattered field or the impulse response due to a monopole source placed inside a closed sphere.
$p(r)$ is the field due to all sources, which represent the reflection of the source within closed domain.
$G(r,r')$ is already defined in Equation~\ref{eq6}, whereas $q(r')$ is the strength of the $j^{th}$ source:

\begin{equation}\label{eq11}
p(r) = \sum_{j}^{N_s} {G(r,r')q(r')}
\end{equation}

Calculating the effect at many target points (e.g., $N_t$) results in quadratic complexity, which can be reduced to $O(N_s + N_t)$ by expanding Equation~\ref{eq11} into a series of spherical harmonics \cite{harmonics1993}, as in Equation~\ref{eq12}. The
non-singular part of the solution can be formulated as in Equation~\ref{eq13}.

\begin{equation}\label{eq12}
p(r) = ik \sum_{n=0}^{\infty} \sum_{m=-n}^{n} S_{n}^{-m}(r_j)R_{n}^{m}(r),  r\leq{r_q}
\end{equation}

\begin{equation}\label{eq13}
p(r) = ik \sum_{n=0}^{\infty} \sum_{m=-n}^{n} C_{n}^{m} R_{n}^{m}(r)
\end{equation}

\begin{equation}\label{eq14}
C_{n}^{m} =\sum_{r_q<R_{max}}Q_q S_{n}^{-m}(r_q)
\end{equation}

In FMM, the coefficients (also known as the multipoles) $C_{n}^{m}$ are computed for each point, and the corresponding series is truncated at $P\ll{N}$, which is the order of expansion, and thus, the complexity is cut down to $P^2$ for a single target. 

\begin{figure}[h]
	\begin{subfigure}{0.49\textwidth}
		\begin{center}
			\includegraphics[scale=1]{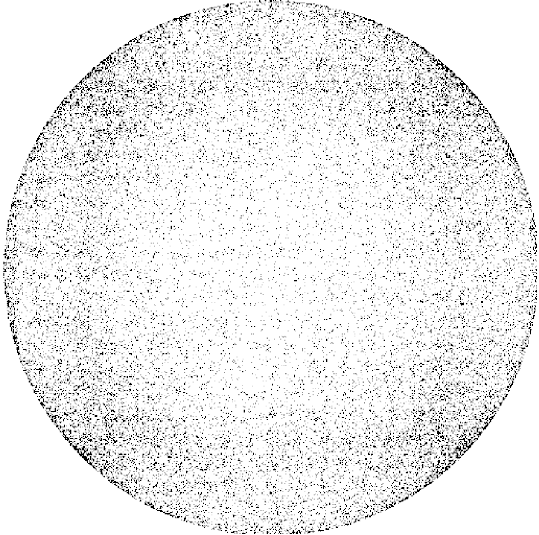}
		\end{center}
		\caption{Original domain before partitioning.}
		\label{fig:sphere_2d}
	\end{subfigure}
	\begin{subfigure}{0.49\textwidth}
		\begin{center}
			\includegraphics[scale=0.4]{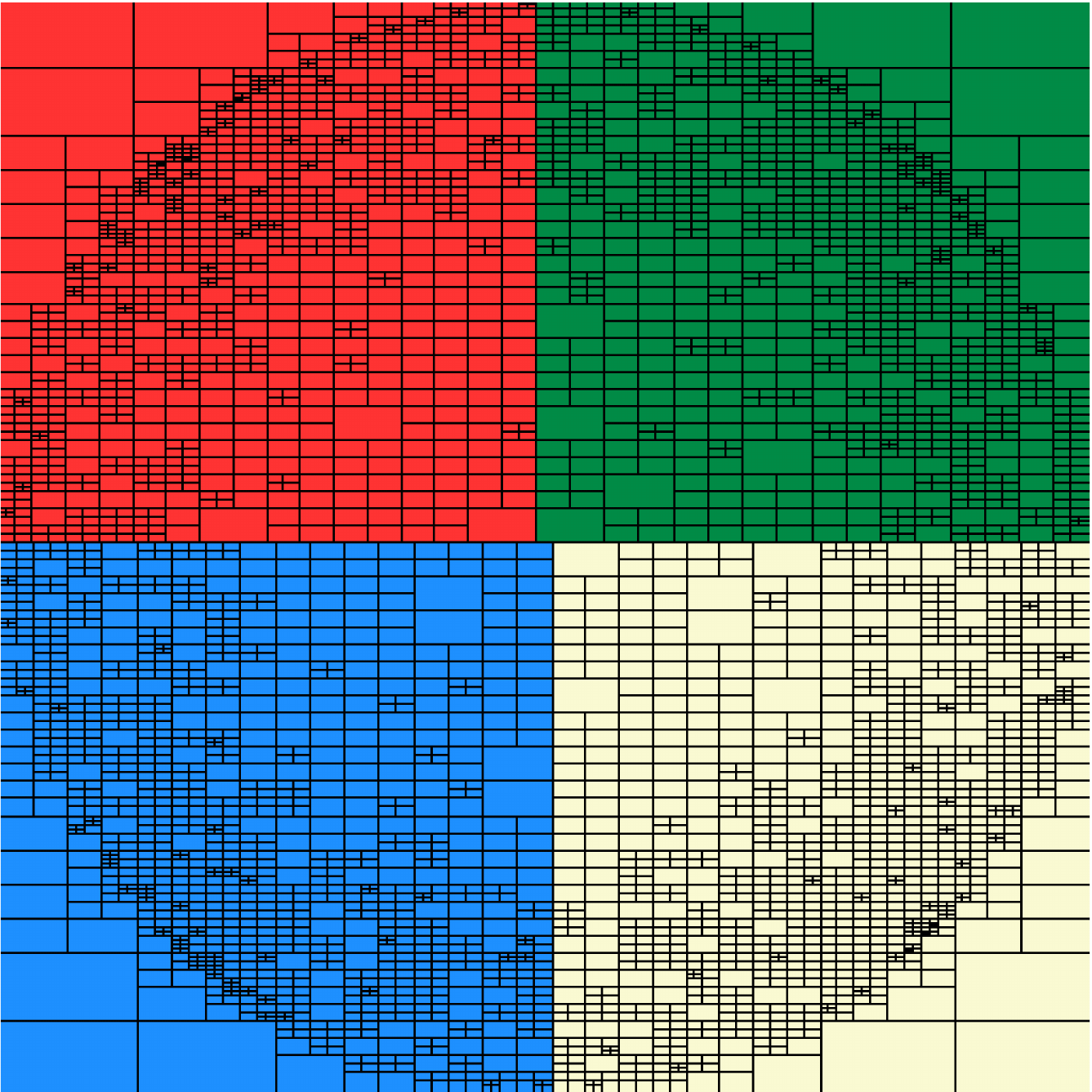}
		\end{center}
		\caption{Quad-tree partitioning on 4 processes.}
		\label{fig:sphere_2d_quad}
	\end{subfigure}\\
		\caption{Particles in a spherical shell 2D domain.}
		\label{fig:quad_tree_part}
\end{figure}

This process is carried out in FMM after hierarchically decomposing the domain into quad/oct tree as shown in Figure~\ref{fig:quad_tree_part}. Then, an upward sweep propagates leaf sources into multipoles. Next, a traverse stage (horizontal pass) is applied to multiply the sources and targets, such that the well-separated cells interact as multipoles, or, otherwise, as particles. A downward pass stage comes afterward to translate local expansions to the corresponding target cell.

%%%%%%%%%%%%%%%%%%%%%%%%%%%%%%%%%%%%%%%%%%%%%%%%%%%%%%%%%%%%%%%%%%%%%%%%%%%%%%%%%%%%%
%%%%%%%%%%%%%%%%%%%%%%%%%%%%%%%%%%%%%%%%%%%%%%%%%%%%%%%%%%%%%%%%%%%%%%%%%%%%%%%%%%%%%

\section{Shared-memory Optimizations}
\label{sec:shared-memory}

In this section, we describe the key performance optimizations
techniques that are applied herein to: 1) improve
the entire application performance, and 2) extract the thread- and data-level parallelisms
within a single compute node.

\subsection{Data-level Parallelism}
\label{sec:data-level_parallelism}

Contemporary processing hardware is equipped with Instruction Set Architecture (ISA) that supports 
Single Instruction, Multiple Data (SIMD) operations on many vectorized data items.
For instance, Intel Xeon Skylake architecture implements two
512-bit Vector Processing Units (VPUs) per core,
by which a single arithmetic instruction can be performed on a large subset of independent, distinct data items.
In the context of our highly optimized FMM-Helmholtz kernels,
we undertake two different vectorization approaches: 1)
we handwrite the vector code for the key kernels using AVX-512 intrinsics, and 2)
we further optimize and fine-tune the kernels to aid the Intel compiler to automatically
generate efficient vector codes.

{\bf Validating the Compiler's Loop Choice:} Treating singularity in the innermost loop of the 3D Helmholtz kernel depicted by Listing~\ref{lst:p2phelmholtz}
through iterating over the high degree Gauss quadrature points within the Particle-to-Particle (P2P) and Source-to-Target
(S2T) routines of FMM involves, in principle, complicated nested {\tt for loops} that involve many conditional statements.
When we analyze the Intel compiler's report of vectorization generated by the
Intel Advisor assistance tool, we find out that the compiler tends to vectorize the innermost loops by default.
However, this does not result in an additional advantage out of the vectorization,
since the spatial and temporal locality of references are well-preserved
when the outer loops are vectorized and strided
\cite{simd_outer_loop, Abduljabbar2017_1, FMM:KNC}.
One way to vectorize such kernels is through populating the 
scalar loop's data using several vector broadcast instructions,
as opposed to fetching a unit stride from a cache line via \texttt{vmov} instruction.
Thus, using \texttt{vbroadcast} instructions imposes lower latency and reciprocal throughput \cite{agner_latency_table}.
Furthermore, lowest latency and highest spatial and temporal locality of reference are achieved with outer loop vectorization, especially in balanced (equal-sized) chucks
inside the nested {\tt for loops}.
In order to construct such vector code, either the compiler needs to be instructed via \texttt{\#pragma simd}, or more aggressively, writing the vector code manually through utilizing intrinsics.
Indeed, both approaches require certain loop optimization techniques (e.g., transformation and unrolling)
\cite{simd_loop_techniques}.
Relying on the compiler to auto-vectorize the code, whenever it is possible, is definitely the right approach and is highly recommended to
guarantee portability and resilience.
Nonetheless, in many cases the compiler could fail to extract the correct or efficient vector code due to assumed data dependencies imposed by the
data structures.

\begin{singlespace}
  \begin{lstlisting}[caption=Trimmed down version of the complex-valued 3D Helmholtz P2P Kernel., captionpos=b, label=lst:p2phelmholtz,float=tp,floatplacement=tbp]
for ( ; i<ni; ++i) {        
  vi_r = real(Bi.SRC);
  vi_i = imag(Bi.SRC);        
  for (j=0; j<nj; ++j) {           
    dX = xi-xj;          
    R2 = norm(dX);
  \\relay self-sigularity to PETSc callback
    if(Bi.PATCH!=Bj.PATCH && R2!=0) {
      real_t R=sqrt(R2);
      if(R<=near_patch_distance) { 
        for (k=0; k<gauss_quad_points; ++k) {  
        \\ near patch singularity treatment                               
        }
      } else {                
        vj_r = real(Bj.SRC);
        vj_i = imag(Bj.SRC);
        src2_r = vi_r*vj_r-vi_i*vj_i;
        src2_i = vi_r*vj_i+vi_i*vj_r;
        invR = 1.0/sqrt(R);
        eikr = 1.0/exp(wave_i*R);
        eikr *= invR;
        eikr_r = cos(wave_r*R)*eikr;
        eikr_i = sin(wave_r*R)*eikr;
        pot_r += src2_r*eikr_r-src2_i*eikr_i;
        pot_i += src2_r*eikr_i+src2_i*eikr_r;
      } 
    }
  }
  Bi.TRG += complex(pot_r, pot_i);
}
\end{lstlisting}
\end{singlespace}

{\bf Writing SIMDizable Code:}
Since some arithmetics are known to be expensive in terms of latency, which could squander
many CPU cycles, modern compilers are designed to avoid such arithmetics as much as possible.
For instance, square root and division operations are very often replaced
by their reciprocal counterpart, whenever the code is compiled with certain optimization flags.
Nevertheless, sometimes the compiler's auto-generated vector codes is suboptimal,
which primarily depends on the scalar source code.
Hence, writing intrinsics seems to be inevitable in such cases.
For example, consider $1/sqrt(R)$ of line 19 in Listing~\ref{lst:p2phelmholtz},
the corresponding assembly code of the Intel compiler comprises of
2 \texttt{vmovups}, 3 \texttt{vmulps}, 1 \texttt{vrsqrt14ps}, 1 \texttt{vfmsub213ps}.
This is fairly a reasonable approach that the compiler adopts to build portable, efficient
vector code.
However, one can write a more efficient code, which uses only \texttt{vmovups} and \texttt{vrsqrt14ps},
via an explicit call to ~\texttt{\_mm512\_rsqrt14\_ps(r)} intrinsic.
On the other hand, a smarter way can achieve both (i.e., efficient vector code generated by the compiler while
avoiding writing explicit SIMD code) through breaking down the
$eikr =(\sqrt{R} \times e^{ikR})^{-1}$ operation to lines 19, 20, and 21 of Listing~\ref{lst:p2phelmholtz}.
Thereby, the compiler automatically understands this transformation, and would extract the most cost-effective, well-optimized
vector code. 

{\bf Optimizing Memory Access:}
It is well-understood that memory bandwidth is a critical obstacle that limits the performance of modern HPC
architectures. As a consequence, one must carefully inspect how cache lines or memory words are fetched
into the vector units, especially since most modern x86 architectures are mounted on dual-socket NUMA nodes, in which
data might physically reside on different address spaces.
Therefore, we develop the FMM core kernels to allocate and reference the particles and tree cells data structures
in the form of Array-of-Structs (AoS).
In addition, AoS enhances the locality of references for interacting particles after they are sorted and indexed based on their Morton order. Cells maintain both indexes and counts of the encapsulated set of particles (see Listing~\ref{lst:str}). This is somehow a simple and compact version of a {\it hash map} associative array abstract data type to map keys to values
(i.e., index and count).
Hence, strided memory access via AVX-512 intrinsics can be utilized to efficiently reference the SRC data structure of
line 2 of Listing~\ref{lst:p2phelmholtz}.
However, having to carry out such low-level manipulation with intrinsics might not be attainable except with a great deal of coding effort,
since it requires manipulation of memory addresses using \texttt{shuffle}, \texttt{permute}, \texttt{gather} and \texttt{scatter} instructions,
which results in non-portable, error-prone, compiler-specific code.
Thus, we code the low-level kernels in such a way that the compiler can extract the most optimal vector code
without the need to explicitly use handwritten intrinsics.

\begin{lstlisting}[label=lst:str, caption= Summary of underlying FMM's data structures., captionpos=b,numbers=none,float=tp,floatplacement=tbp]
struct particle_t {
  num SRC;
  num COORD[3];
} __attribute__((aligned (64)));

struct cell_t {
  particle_t* b_ptr;
  size_t b_count;
} __attribute__((aligned (64)));

particle_t* particles;
cell_t*     cells;
\end{lstlisting}

\subsection{Thread-level Parallelism}
\label{sec:thread-level_parallelism}

\begin{figure}[h]
    \begin{subfigure}{0.49\textwidth}
      \begin{center}
        \includegraphics[scale=0.75]{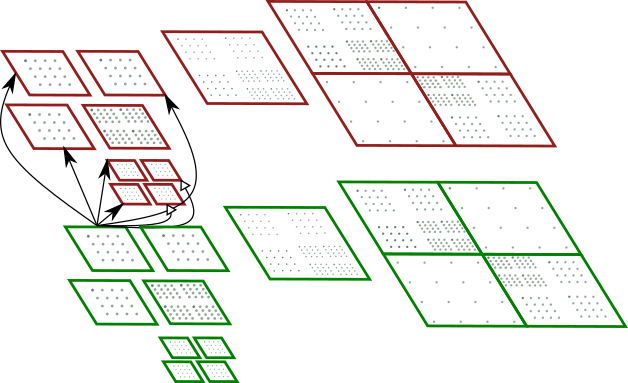}
      \end{center}
    \caption{Dual-tree traversal.}
    \label{fig:dtt}
    \end{subfigure}
    \begin{subfigure}{0.49\textwidth}
      \begin{center}
        \includegraphics[scale=0.95]{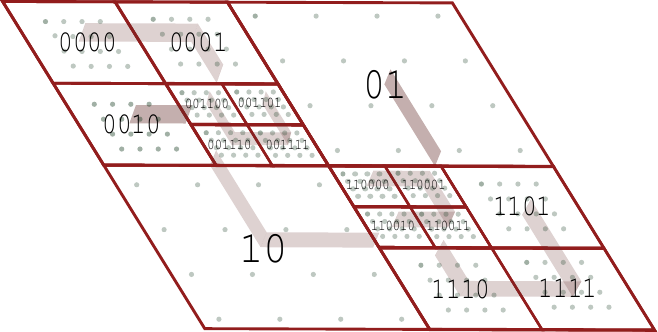}
      \end{center}
    \caption{Warren and Salmon original FMM.}
  \label{fig:lbt}
    \end{subfigure}\\

  \caption{Geometric FMM tree partitioning.}
  \label{fig:scalability}
\end{figure}

The well-known linear complexity of FMM is accomplished by mapping positions of tree cells into Morton/Hilbert keys, and probing each cell for its interaction list by interpolation of bits while traversal \cite{Warren2013}.
Figure~\ref{fig:lbt} depicts geometric tree partitioning of cells, which map positions to binary keys. The line segments represent the original level-wise Hilbert orders. However,
%especially for Laplace and Cosmological applications,
maximizing thread-level parallelism has been proven more successful using the Dual Tree Traversal (DTT) approach \cite{Yokota2013a,Lange2014}, which is known for its adaptability to multi- and many-core emerging architectures. For example, to find the scattered field at target positions encoded by \texttt{0000} Hilbert order in Figure~\ref{fig:lbt}, DTT simultaneously traverses source and target trees, and recursively uncovers the cell-cell interaction list (see Figure~\ref{fig:dtt}).

One of the disadvantages of DTT is its sensitivity to granularity and size of tasks. We believe that it is crucial to consider maximizing locality of reference, by which we can obtain many advantages of high concurrency and throughput in latency-bound compute kernels. Equation~\ref{eq:th_performance_model} models the optimal parameters (i.e., ($s,c$)) to fill the L2/L3 caches, and maximize concurrency and throughput as well as cache locality of reference. The multiplier ``$2$'' is inclusive of source and target, $csize$/$bsize$ is the cell/body structure size in bytes,
$s$ is the task spawning parameter,
$c$ is the number of bodies per leaf cell,
$\log{\frac{s}{c}}$ is the depth of recursive branch, and L2/L3 Last Level Cache (LLC) size in bytes.

\begin{equation}\label{eq:th_performance_model}
\begin{aligned}
 \min_{s,c} f(s,c) 	&= (M2L_{size} + P2P_{size}) - \text{L2/L3 Cache} \\
 										&= 2 \times  \text{ c } \times  \text{ task size } \times \text{ nthreads/core }\\
										&\times [(csize \times \log{\frac{s}{c}}) + bsize] - \text{L2/L3 Cache}\\
\end{aligned}
\end{equation}

%%%%%%%%%%%%%%%%%%%%%%%%%%%%%%%%%%%%%%%%%%%%%%%%%%%%%%%%%%%%%%%%%%%%%%%%%%%%%%%%%%%%%
%%%%%%%%%%%%%%%%%%%%%%%%%%%%%%%%%%%%%%%%%%%%%%%%%%%%%%%%%%%%%%%%%%%%%%%%%%%%%%%%%%%%%

\section{Distributed-memory Optimizations}
\label{sec:distributed-memory}

Having explained the key shared-memory optimization techniques to improve
the application's single node performance, we elaborate on the
main distributed-memory, MPI-based implementation aspects, which focus upon
building robust mechanisms for extreme scale
partitioning, load balancing, and communication reducing.

\subsection{Partitioning and Load Balancing}
\label{subsec:partition_load_balancing}

Despite the fact that FMM is asymptotically linear in terms of the theoretical time complexity, a naive distributed-memory workload partitioning can
take away the advantages of using an optimal algorithm.
In general, For $N$-body codes, the domain has to be decomposed in order to maximize local computations (near-field), while minimizing the volume of global communications. Additionally, the computation and communication ratios are balanced in such a way that large amounts of
computation can be carried out between communication events.

{\bf Pre-partitioning Stage:} One of the obstacles that we run into when doing large-scale experiments is reading large input meshes from persistent file systems (e.g., \texttt{LUSTRE}).
We handle a variety of formatted mesh ASCII files, including \texttt{GMSH} and \texttt{IDEAs}). Also, we develop a partitioning scheme that is specifically designed to work well with our FMM implementation.
Hence, a third-party partitioning tool (e.g., \texttt{ParMETIS}) can be used in our
application code.
To partition the file, we implement a straightforward approach, which works well up
to small number of nodes (i.e., up to 64 compute nodes). This approach reads the mesh file on the parent compute node (MPI rank 0),
and then partitions and broadcasts the data to the rest of the ranks.
However, due to memory and network limitations, this approach is not scalable.
Therefore, we develop an external C++ routine that builds an intermediate binary 
big-endian mesh file, in which we include extra characteristics related to the mesh
distribution.
We abstract out the information about nodes and elements of the underlying mesh format,
and translate them into coordinates corresponding to the elements.
Each MPI process, thereafter, calculates its offset and reads a block of the data
in a collision-free manner prior to the domain-specific partitioning.
Consequently, we conserve a great deal of computational time that is consumed to
parse TeraBytes of ASCII numbers.
Furthermore, this approach does not require synchronization of file reads or even using MPI I/O.
We simultaneously load balance the network reads in terms of bandwidth and throughput.

{\bf Partitioning Stage:}
The core building block of our partitioning implementation for large-scale is the 
the modified Orthogonal Recursive Bisection (ORB) of \cite{Abduljabbar2017_2}.
Since the growth of interaction lists is governed by
distribution randomness, workload size, and communication volume cost,
obtaining an optimal partitioning is extremely challenging.
For example, cells have different sizes of interaction lists, and therefore,
equidistributing the results of the ORB might lead to a suboptimal balance of the workload.
Hence, we further improve the efficiency our partitioning strategy
through weighting by workload.
In other words, we use the workload size of the previous time step to weight the particles, so
that we promote a cost-effective and adaptive load balance across the MPI ranks.
Originally, this technique is implemented in the original Hashed Oct-Tree (HOT) \cite{Warren2013}, and we purposely shape and fine-tune it to be
applicable and coherent to our application code.
One drawback of the weighting scheme is that it only balances the workload, but not the communication volume.
Therefore, there have been significant efforts to employ
a graph partitioning tool to use the workload as node-weights and communication as edge-weights.
It aims to create partitions that maintain both an optimal balance of the workload as well as the communication \cite{Cruz2011}.
Furthermore, this method has only been compared with Morton key splitting without weights.
Hence, we balance the workload and communication simultaneously by calculating the weight for the $i^{th}$ particle $w_i$ according Equation~\ref{eq:weight},
where $l_i$ is the local interaction size, $r_i$ is the remote interaction list size, and $\alpha$ is a constant that is optimized over the time steps to minimize the total runtime. $l_i+r_i$ is the total interaction list size and represents the workload, while $r_i$ reflects the amount of communication. By adjusting the coefficient $\alpha$, one can amplify/damp the importance of communication balance. Making this an optimization problem to minimize the total runtime is what we prefer over minimizing the load imbalance since the latter is not our final objective. 
In addition, the variables $l_i$ and $r_i$, and the total runtime are already measured in the present code, so the information is available with negligible cost.

\begin{equation}
w_i=l_i+\alpha*r_i
\label{eq:weight}
\end{equation}

\begin{figure}[h]
  \centering
  \includegraphics[scale=0.75]{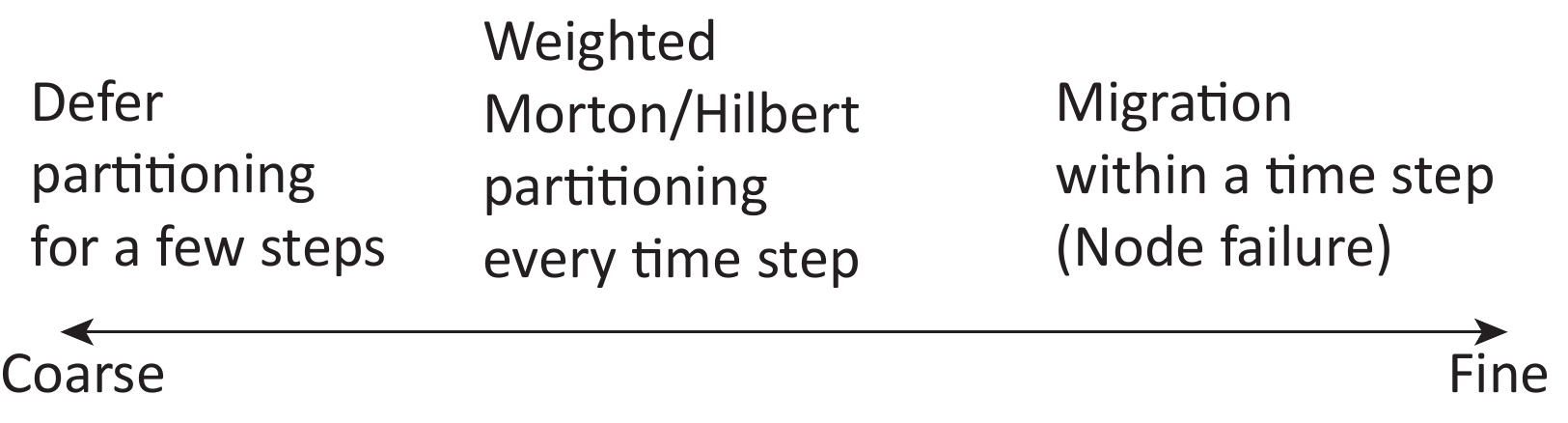}
  \caption{Granularity of partitioning.}
  \label{fig:orb_weighting}
\end{figure}

Figure~\ref{fig:orb_weighting} manifests the granularity spectrum for the partitioning phase.
We restrict the partitioning phase to the granularity of a configurable steps (GMRES restart), because it is a too costly process for a stationary system.
Also, we have a limited need for partitioning at fine granularity.
We instantiate an artificial FMM partitioning and traverse call before invoking the GMRES solver,
in order to apply weighting based on both workload as well as communication (see Equation~\ref{eq:weight}).

\subsection{Communication Reduction}
\label{subsec:communication-reduction}

Once we determine the multipole expansions for every local cell,
we pass the multipole expansions to the necessary processes
in a ``sender-initiated'' fashion \cite{Dubinski1996}.
This reduces the latency, since we communicate only once
rather than sending a request to remote processes and then receiving the data.
Such ``sender-initiated'' communication schemes
are common in cosmological $N$-body codes,
since they tend to use only monopoles.

\begin{figure}[h]
  \centering
  \includegraphics[scale=0.28]{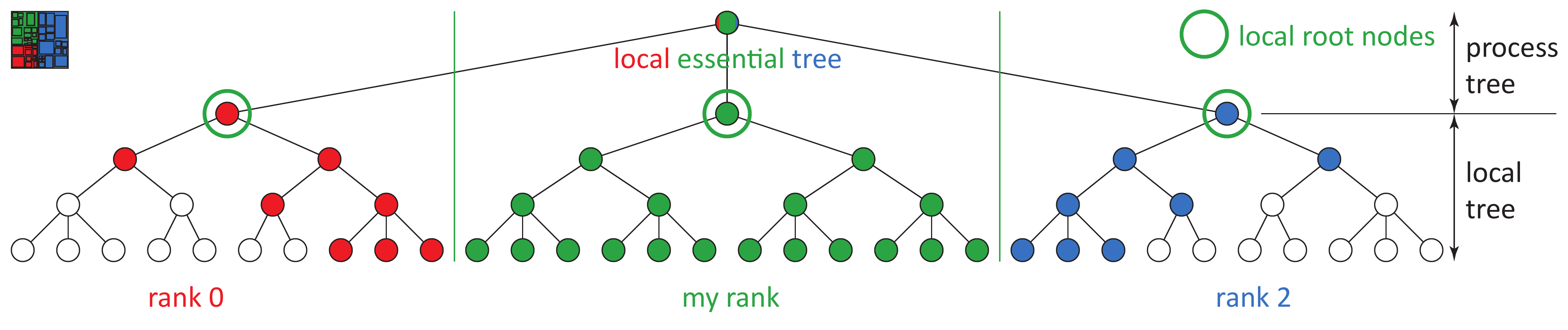}
  \caption{How LET is grafted.}
  \label{fig:tree}
\end{figure}

Figure~\ref{fig:tree} presents the Local Essential Tree (LET) that is formed from the information sent from the remote processes by simply grafting the root nodes of the remote trees.
In a conventional parallel FMM code, a global octree is formed and partitioned using either HOT or ORB.
Therefore, the tree structure is severed in many places, which complicates the merging of the LET.
The merging LET code consumes a large bulk of the total execution time of FMM, and thus,
incorporating additional features (e.g., periodic boundary conditions, mutual interaction, more efficient translation stencils, and dual tree traversals) would affect the runtime.
However, we geometrically separate the global tree structure from the local tree structure
to merge the tree in a a single time step as shown by Figure~\ref{fig:tree}, and are able to incorporate extended features.

\begin{figure}[h]
  \centering
  \includegraphics[scale=0.48]{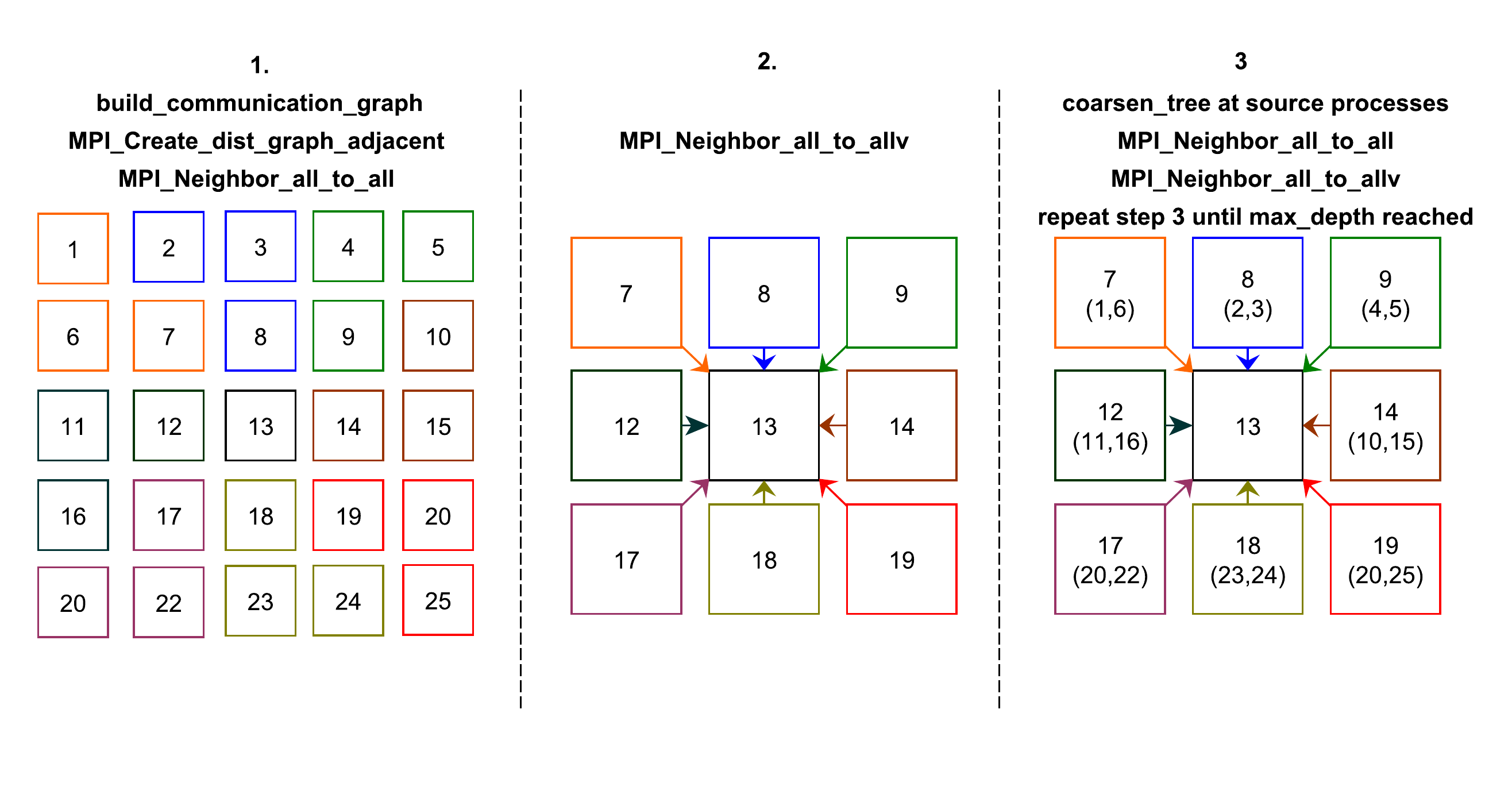}
  \caption{$\mathcal{HSDX}$: A neighborhood-based communication protocol.}
  \label{fig:hsdx}
\end{figure}

While the remote information for the LET is being transferred, the local tree can be traversed. Conventional fast $N$-body methods overlap the entire LET communication with the entire local tree traversal. The LET communication becomes a bulk-synchronous \texttt{MPI\_alltoallv} type communication. To achieve scalability to all of Shaheen, we embrace the neighborhood-based communication protocol of \cite{Abduljabbar2017_2}, namely  Hierarchical Sparse Data eXchange ($\mathcal{HSDX}$) (see Figure~\ref{fig:hsdx}).
The algorithm unfolds into 3 stages, as follows:

\begin{enumerate}[noitemsep, leftmargin=*]
  \item
  The MPI communication graph maps the ranks to adjacent nodes by calling \texttt{MPI\_Create\_dist\_graph\_adjacent}.
  This mapping uses their logical near counterparts from the global tree of Figure~\ref{fig:tree}, and each color indicates a single communication route for the LET
  (e.g., rank 13 in Figure~\ref{fig:hsdx}).
  \item
  Perform communication of the neighboring trees/ranks.
  \item
  Communicate the neighbors of neighbors' data multiple times, since they are already available at adjacent nodes from previous communication steps.
\end{enumerate}

%%%%%%%%%%%%%%%%%%%%%%%%%%%%%%%%%%%%%%%%%%%%%%%%%%%%%%%%%%%%%%%%%%%%%%%%%%%%%%%%%%%%%
%%%%%%%%%%%%%%%%%%%%%%%%%%%%%%%%%%%%%%%%%%%%%%%%%%%%%%%%%%%%%%%%%%%%%%%%%%%%%%%%%%%%%

\section{Experimental Setup and Workload Characterization}
\label{sec:Experiments}

This section describes our experimental platforms, the
datasets, and the scientific performance engineering methodologies
that are used to analyze and present the performance evaluation results.

\subsection{Software Stack and Hardware Configuration}
\label{subsec:hard_soft}

The source code is written in C$++$.
We use PETSc release version 3.8 built on top of
Intel Parallel Studio version 2018 Update 1, which includes
Intel C/C++ compiler, Threading Building Blocks (TBB), OpenMP, Cilk Plus\footnote{
Intel Cilk Plus is being deprecated in the 2018 release of Intel Software Development Tools.
}, MPI, and Math Kernel Library (MKL).
PETSc scalar type is set to complex. PETSc is complied with the C$++$ compiler, and
the FORTRAN kernels are set to generic for faster complex number performance.
All experiments are performed with the {\tt -O3} compiler optimization flag,
and OpenMP affinity is set to scatter via {\tt KMP\_AFFINITY=scatter}.
The pinning and binding of the thread contexts and the MPI ranks are set to
target a specific {\tt quadrant/tile/core} on KNL, and a specific socket/core on CPU.
Furthermore, we use {\tt numactl} Linux command to control binding and interleaving of memory channels.
Table~\ref{tbl:specs} summarizes the specifications of the Intel x86 architectures considered herein.

\begin{table}[h]
  \centering
  \caption{Hardware specifications.}
  \label{tbl:specs}
  \begin{tabular}{l|l|l|l|}
    \cline{2-4}
                                            & \multicolumn{1}{c|}{KNL}  & \multicolumn{1}{c|}{Haswell}  & \multicolumn{1}{c|}{Skylake}  \\ \hline
    \multicolumn{1}{|l|}{Family}            & x200                      & E5V3                          & Scalable                      \\ \hline
    \multicolumn{1}{|l|}{Model}             & 7290                      & 2670                          & 8176                          \\ \hline
    \multicolumn{1}{|l|}{Socket(s)}         & 1                         & 2                             & 2                             \\ \hline
    \multicolumn{1}{|l|}{Cores}             & 72                        & 32                            & 56                            \\ \hline
    \multicolumn{1}{|l|}{GHz}               & 1.50                      & 2.60                          & 2.10                          \\ \hline
    \multicolumn{1}{|l|}{Watts/socket}      & 245                       & 120                           & 165                           \\ \hline
    \multicolumn{1}{|l|}{DDR4 (GB)}         & 192                       & 128                           & 264                           \\ \hline
    \multicolumn{1}{|l|}{Frequency Driver}  & acpi-cpufreq              & acpi-cpufreq                  & acpi-cpufreq                  \\ \hline
    \multicolumn{1}{|l|}{Max GHz}           & 1.50                      & 2.60                          & 2.10                          \\ \hline
    \multicolumn{1}{|l|}{Governor}          & conservative              & performance                   & ondemand                      \\ \hline
    \multicolumn{1}{|l|}{Turbo Boost}       & $\checkmark$              & $\checkmark$                  & $\checkmark$                  \\ \hline
  \end{tabular}
\end{table}

For the large-scale experiments, we use KAUST's Shaheen XC40, the rank 20 supercomputer according to the TOP500 list of November 2017.
The system consists of 6,174 compute nodes,
each of which is equipped with a dual socket Intel Haswell CPU (see Table~\ref{tbl:specs}).
The entire system has 197,568 hardware cores, and 786 TB of main memory.
(Note that in our experimentations we consider the full scale of Shaheen is 6,144 compute nodes with
196,608 hardware cores, by which we purposely leave 30 compute nodes (i.e., 960 hardware cores) untouched for logistical configurations.)
The compute nodes are connected by the Cray Aries interconnect with dragonfly topology,
which provides for a maximum of 3 hops for a message between any pair of nodes.
Theoretically, Shaheen has a peak double precision floating point performance of 7.2 PFlop/s.

\subsection{Dataset Description}
\label{subsec:dataset}

\begin{figure}[h]
  \begin{center}
    \includegraphics[scale=0.2]{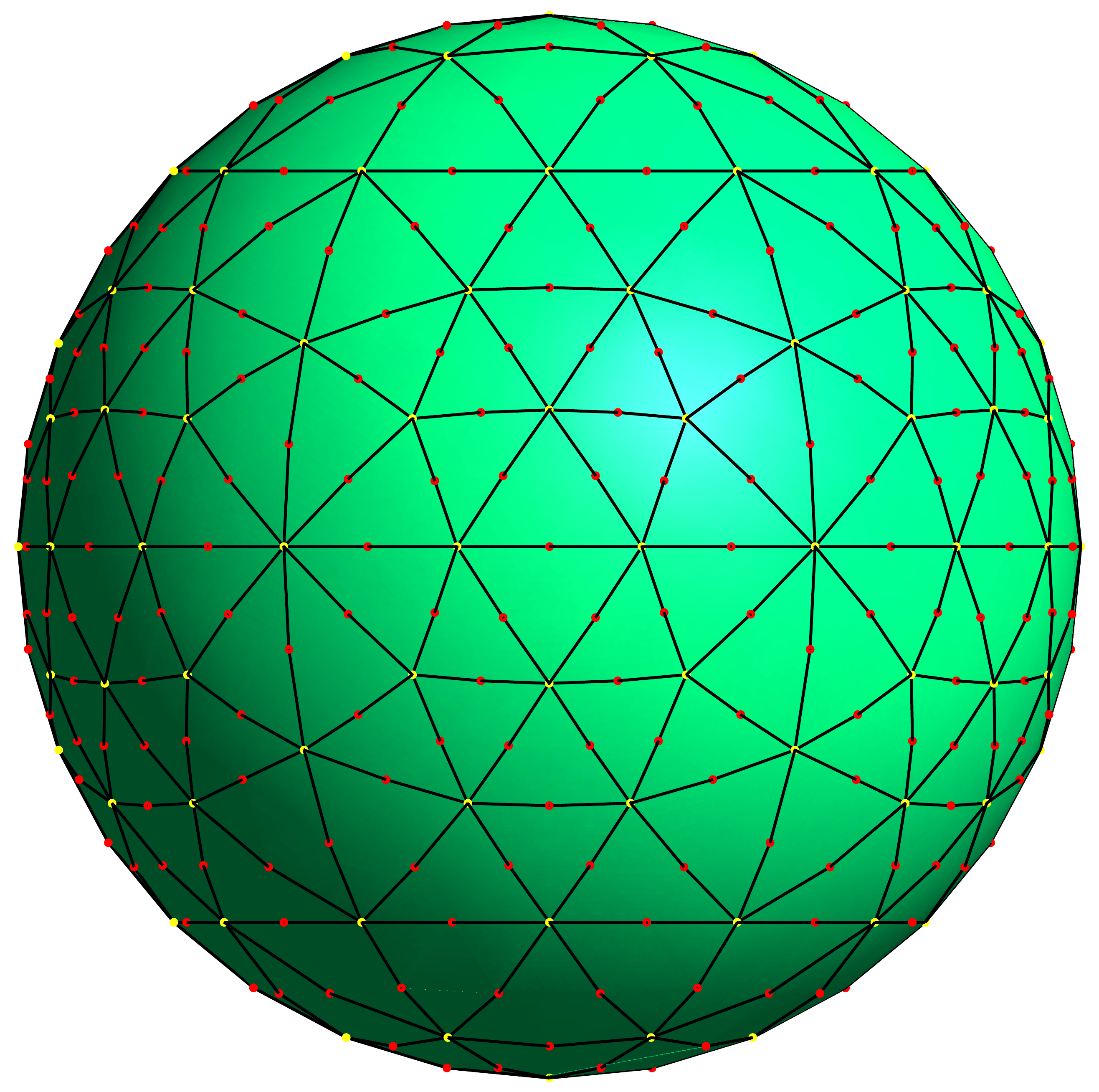}
  \end{center}
  \caption{Example of a curved-triangle mesh.}
  \label{fig:mesh}
\end{figure}

In this paper, we consider a spherical object to scatter an incident uniform plane wave. 
The sphere's radius is set to $a=1m$, and the medium speed of sound is set to $343m/s$.
We use a high-order curvilinear (curved-triangle) meshes, in which
every mesh element (triangle) has 6 quadrature points (unknowns/Degrees-of-Freedom (DoF)):
3 points associated with the triangle nodes and 3 points associated
with the triangle edges (see Figure~\ref{fig:mesh}).
The meshes are generated via IDEAS\textsuperscript{\sffamily\textregistered}.
For detailed mesh properties, Table~\ref{tbl:dataset} describes the specifications of our experimental datasets.

\begin{table}[h]
\centering
  \caption{Mesh dataset descriptions.}
  \label{tbl:dataset}
  \begin{tabular}{|c|c|c|c|c|}
    \hline
     Mesh  & Elements    & Nodes       & Edges           & Number of unknowns (N) \\ \hline
   A     & 156         & 312          & 468          & 936               \\ \hline
   B     & 3,156       & 6,312        & 9,468        & 18,936            \\ \hline
   C     & 7,274       & 14,548       & 21,822       & 43,644             \\ \hline
   D     & 14,338      & 28,676       & 43,014       & 86,028             \\ \hline
   E     & 22,370      & 44,740       & 67,110       & 134,220            \\ \hline
   F     & 41,258      & 82,516       & 123,774      & 247,548             \\ \hline
   G     & 60,204      & 120,408      & 180,612      & 361,224             \\ \hline
   H     & 93,590      & 187,180      & 280,770      & 561,540             \\ \hline
   I     & 115,454     & 230,908      & 346,362      & 692,724             \\ \hline
   J     & 159,288     & 318,576      & 477,864      & 955,728             \\ \hline
   K     & 250,514     & 501,028      & 751,542      & 1,503,084           \\ \hline
   L     & 314,212     & 628,424      & 942,636      & 1,885,272           \\ \hline
   M     & 374,360     & 748,720      & 1,123,080    & 2,246,160             \\ \hline
   N     & 1,497,440   & 2,994,880    & 4,492,320    & 8,984,640             \\ \hline
   O     & 5,989,760   & 11,979,520   & 17,969,280   & 35,938,560             \\ \hline
   P     & 23,959,040  & 47,918,080   & 71,877,120   & 143,754,240            \\ \hline
   Q     & 95,836,160  & 191,672,320  & 287,508,480  & 575,016,960             \\ \hline
   R     & 383,344,640 & 766,689,280  & 1,150,033,920  & 2,300,067,840             \\ \hline
  \end{tabular}
\end{table}

\subsection{Experimental Setup}
\label{subsec:setup}

To report the most accurate performance measurements irrespective of the hardware states and conditions,
we apply several state-of-the-practice scientific performance engineering methodologies, which
overcome any possible hardware-oriented performance variations.

The reported runtime results are summarized using the
arithmetic mean across multiple independent runs, which form the sample space.
The reported floating point rates, on the other hand, are summarized using the
harmonic mean \cite{Hoefler:2015:SBP:2807591.2807644}.
Unless otherwise reported, we average approximately 50 runs for every experiment, except
for the large-scale results, in which we are constrained by the available core hours.
Thus, we reduce the size of the sample space for every run based on the
available core hours (i.e., we roughly average between 5 to 10 runs for every experiments, based
upon the problem size and the wall-clock time of a specific run).
In addition, an error bar is drawn to show the +/- standard deviation of the mean for each experimental sample.

%%%%%%%%%%%%%%%%%%%%%%%%%%%%%%%%%%%%%%%%%%%%%%%%%%%%%%%%%%%%%%%%%%%%%%%%%%%%%%%%%%%%%
%%%%%%%%%%%%%%%%%%%%%%%%%%%%%%%%%%%%%%%%%%%%%%%%%%%%%%%%%%%%%%%%%%%%%%%%%%%%%%%%%%%%%

\section{Performance Evaluation Results}
\label{sec:results}

We describe in this section the single- as well as the multi-node
performance evaluation results for the acoustics application.

\subsection{Data-level Parallelism Results}
\label{subsec:perf-data-level}

\begin{figure}[h]
  \begin{minipage}{0.49\textwidth}
    \begin{center}
      \includegraphics[scale=0.44]{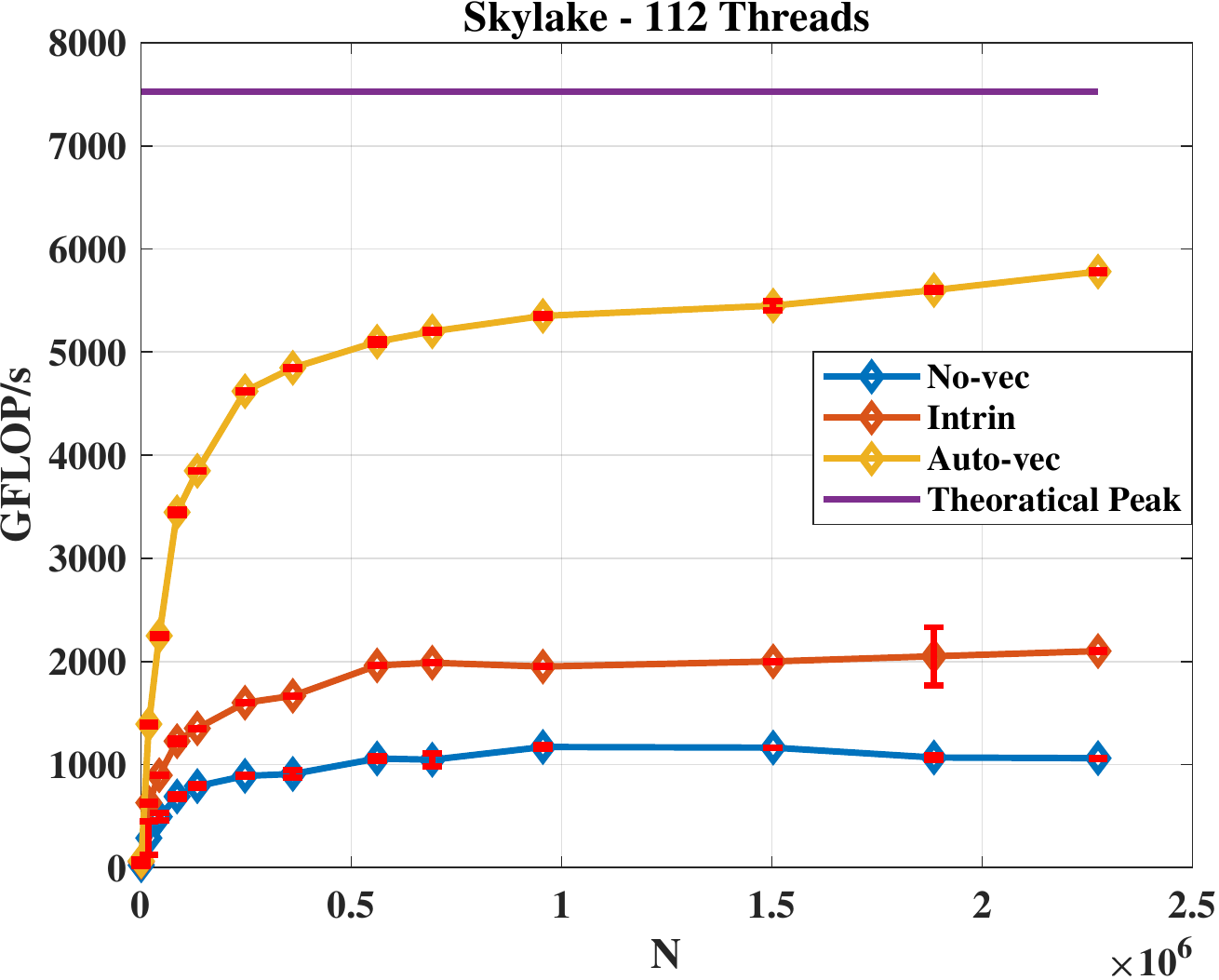}
    \end{center}
  \end{minipage}
  \begin{minipage}{0.49\textwidth}
    \begin{center}
      \includegraphics[scale=0.44]{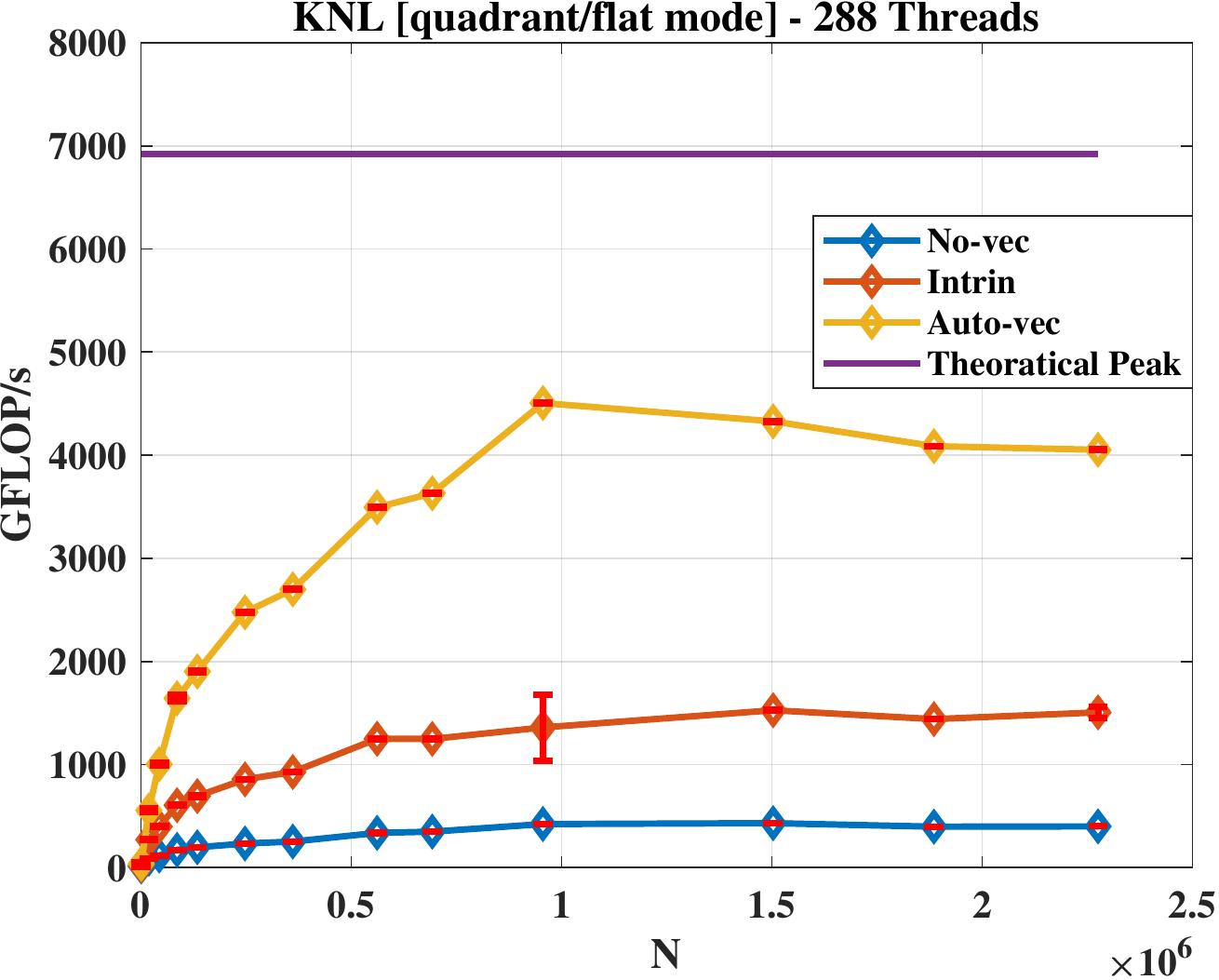}
    \end{center}
  \end{minipage}

  \caption{Floating point performance of the P2P Helmholtz kernel with different vectorization means.}
  \label{fig:flopsP2P}
\end{figure}

The most floating point intensive portion of our application code is the matrix-vector multiplication kernel, which is accelerated by FMM.
To this end, speeding up such kernel increases the entire application's performance, thus maximizing the performance at extreme scale.
Figure~\ref{fig:flopsP2P} shows the single precision floating point performance of the Helmholtz P2P
kernel on both Intel KNL as well as Skylake across different meshes. Relying on the Intel compiler's auto-vectorization, the best performance we achieve on a single node of Skylake is roughly 5.7 TFLOP/s. Since the peak single precision floating point performance of our used edition of Skylake is roughly 7.5 TFLOP/s,
our optimized P2P kernel obtains 77\% out of the peak.
Handwritten vectorization via AVX-512 intrinsics, on the other hand, achieves at most 27\% (i.e., approximately 2.0 TFLOP/s).
In contrast, the non-vectorized version gains a maximum of 14\% (i.e., roughly 1.0 TFLOP/s), which means that our optimized version
maintains an average 5.4x speedup relative to the scalar code.
We observe a similar behavior on KNL, where the auto-vectorization achieves at most 60\% (i.e., approximately 4.5 TFLOP/s) of KNL's single
precision peak floating point performance (i.e., roughly 6.9 TFLOP/s). This is 10x speedup, on average, compared to the scalar code (roughly 0.43 TFLOP/s).

In the handwritten AVX-512 intrinsics code, we explicitly reference
the auxiliary fields data structure ({\tt cell\_t}) that includes the coordinates, source, and target values using {\tt\_mm512\_setr\_ps} intrinsic.
The explicit setting of the vector register values is
carried out since the ({\tt cell\_t}) data structure is allocated as an Array-of-Structs (AoS),
which is a well-known harmful approach to the code vectorization.
Hence, the explicit vectorization does not improve much over the scalar code.
In addition, the compiler initially fails to auto-vectorize the code as a consequence of the assumed
data dependency resulting from the Array-of-Structs allocation.
One solution to this is to change the struct allocation of the code \cite{CFDmanycore}. However, this
can be a daunting proposition, in which the whole ExaFMM code needs to be adjusted to
be compatible.
This would also result in a significant performance reduction due to the loss of cache locality of references, an advantage granted by using AoS in conjunction with Morton orders.
The other approach followed here, as explained earlier in the paper, is to
simplify the P2P kernel to allow the compiler to generate an efficient vector code.
In this approach, we rely on the compiler to find an efficient way to deal with the
hurdles of the AoS allocation.
As a result of such a kernel rewrite, the compiler manages to generate an efficient code that achieves roughly 77\% out Skylake's peak, and nearly 85\% out of the Intel MKL SGEMM\footnote{
  SGEMM is the Single Precision real valued GEneral Matrix-Matrix Multiplication kernel.
} performance on Skylake (i.e., 6.7 TFLOP/s).
To verify how AoS data is loaded, we scrutinize the generated assembly vector code, and we find out that the compiler
performs multiple vector strided loads with a stride size equal to the size of the SIMD lane. This approach efficiently deals with the AoS allocation, but it requires very complex assembly and intrinsics coding \cite{CFDmanycore}.
Thus, the fact that the compiler manages to generate such code, indeed with the help of our subtle kernel simplifications,
saves time and effort. Consequently, it protects against writing error-prone and non-portable kernel codes.

\subsection{Thread-level Parallelism Results}
\label{subsec:perf-thread-level}

\begin{figure}[h]
  \begin{minipage}{0.49\textwidth}
    \begin{center}
      \includegraphics[scale=0.43]{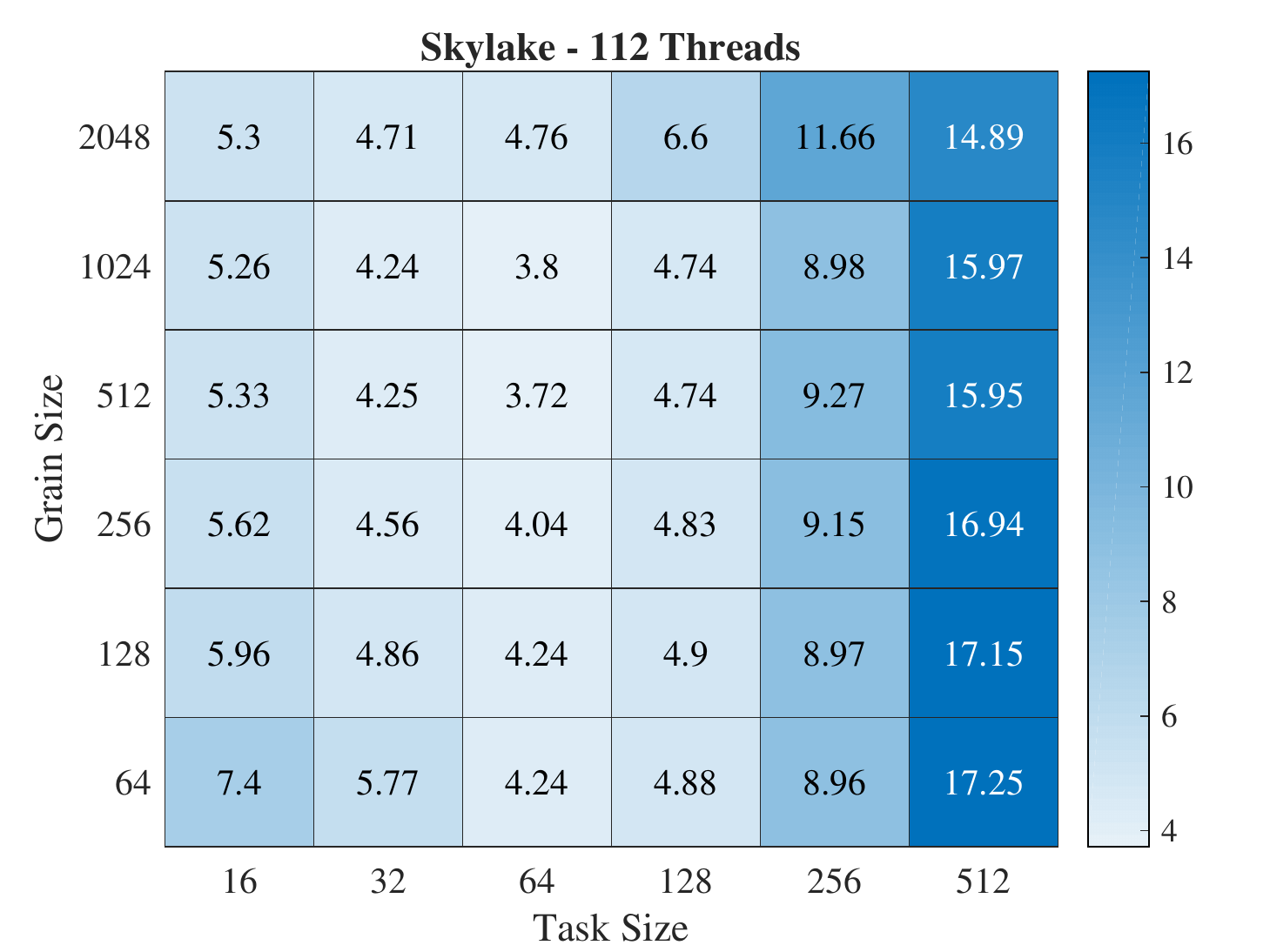}
    \end{center}
  \end{minipage}
  \begin{minipage}{0.49\textwidth}
    \begin{center}
      \includegraphics[scale=0.43]{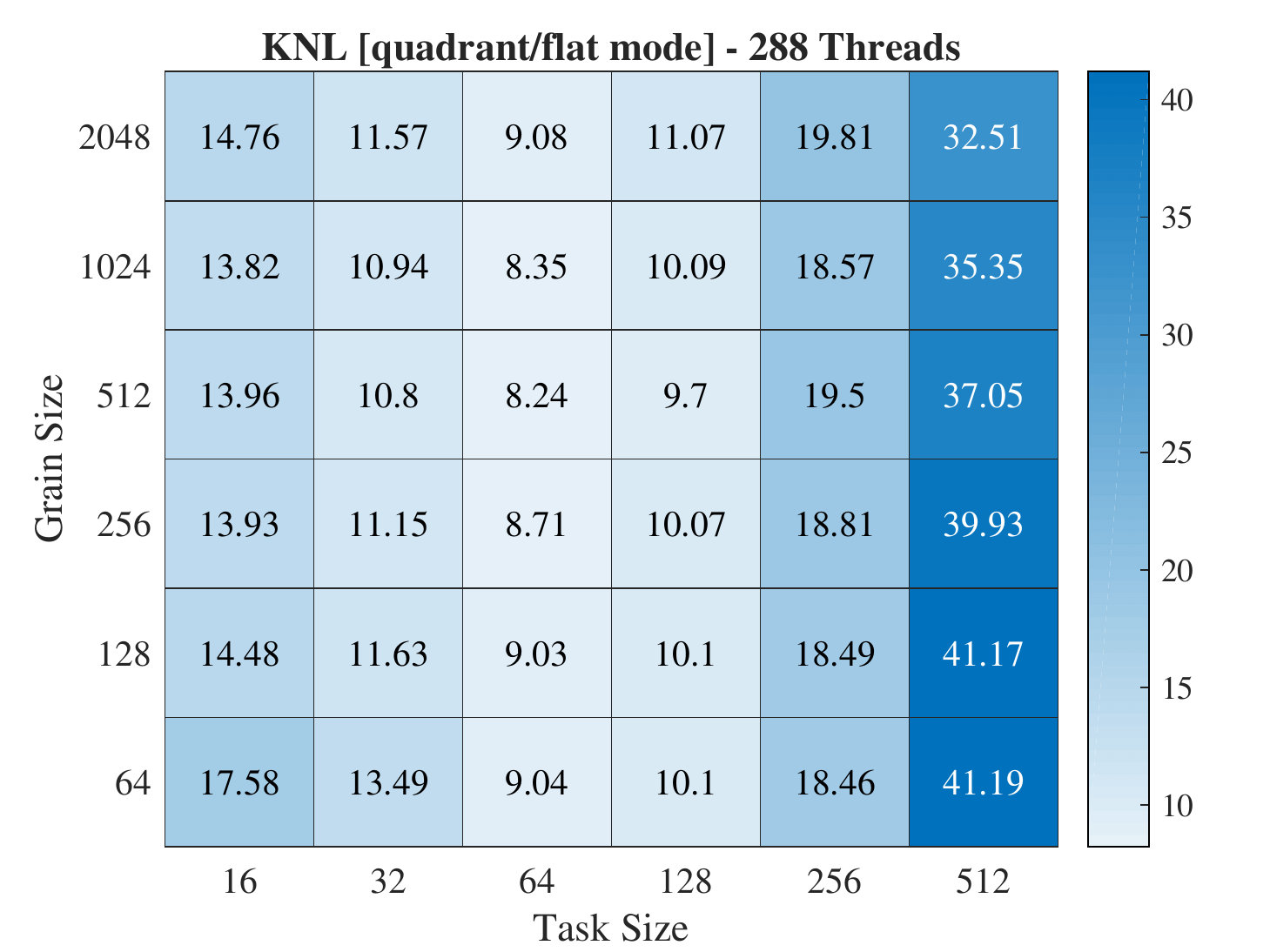}
    \end{center}
  \end{minipage}

  \caption{Tuning the threading parameters performance of the traversal Helmholtz kernel. Performance is normalized by the arithmetic mean of the runtime.}
  \label{fig:heat_map0}
\end{figure}

Figure~\ref{fig:heat_map0} shows the runtime performance of the traversal Helmholtz kernel on KNL and Skylake with varying the grain ($y$-axis) and cell sizes ($x$-axis). 
If the grain size ($x$) is less than or equal to the summation of bodies enclosed within the source and target,
then a task will be spawned by the work scheduler. Essentially, this means having smaller grain sizes would fork more threads,
which implies creating a fine-grained thread pool. However, as manifested from Figure~\ref{fig:heat_map0}, having a coarse-grained
thread pool with a smaller number of tasks, the performance improves.
Similarly, having a smaller size of a task ($y$), which is the number of bodies within a leaf-cell, exhibits better performance.
Thus, it is important to think of processing not so much in terms of long functional threads
but smaller-sized tasks that can be handed to a thread pool.
In addition, a careful number of generated tasks must be considered to avoid overfilling the scheduler thread pool.

\begin{figure}[h]

  \begin{minipage}{0.49\textwidth}
    \begin{center}
      \includegraphics[scale=0.43]{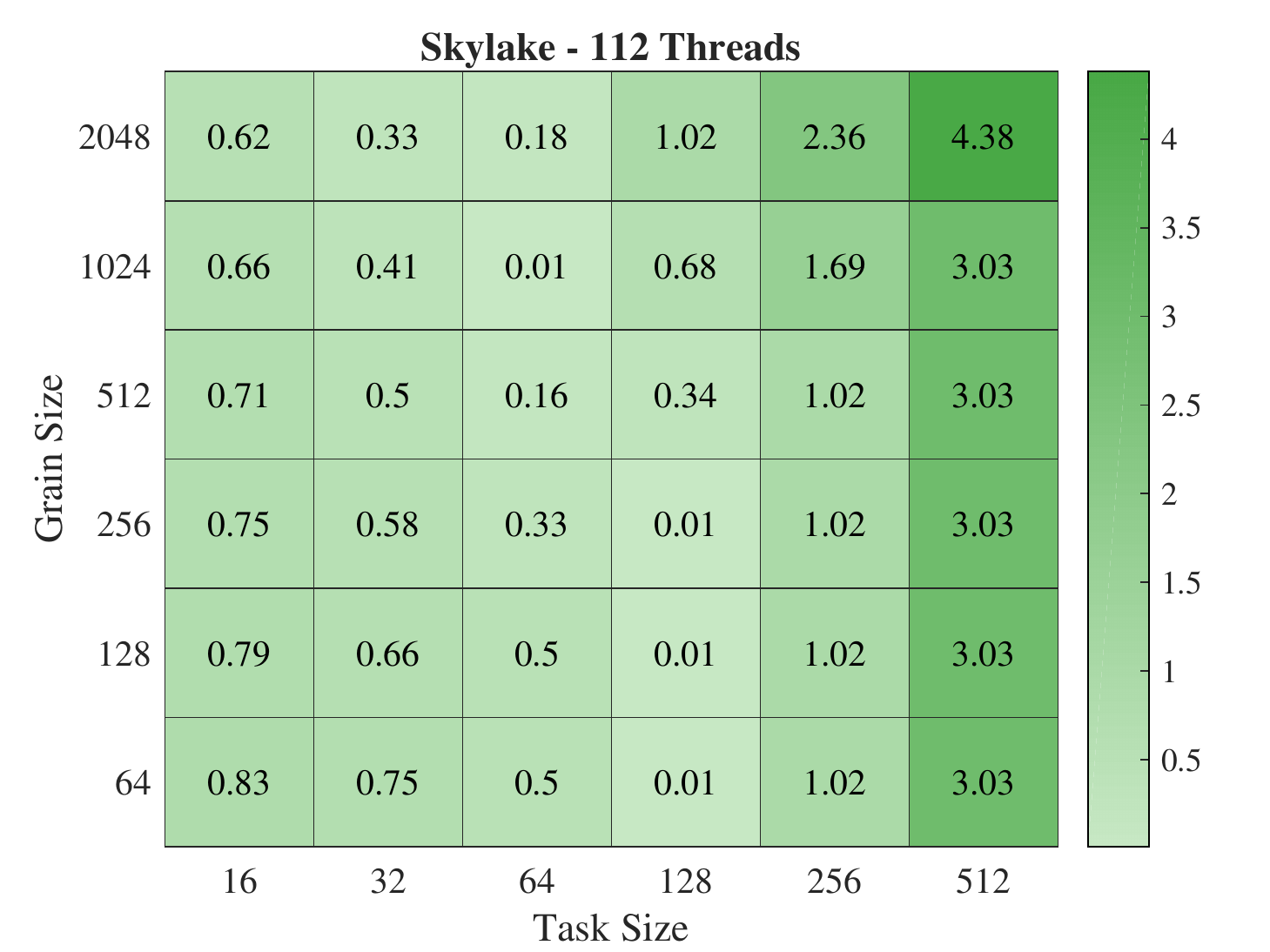}
    \end{center}
  \end{minipage}
  \begin{minipage}{0.49\textwidth}
    \begin{center}
      \includegraphics[scale=0.43]{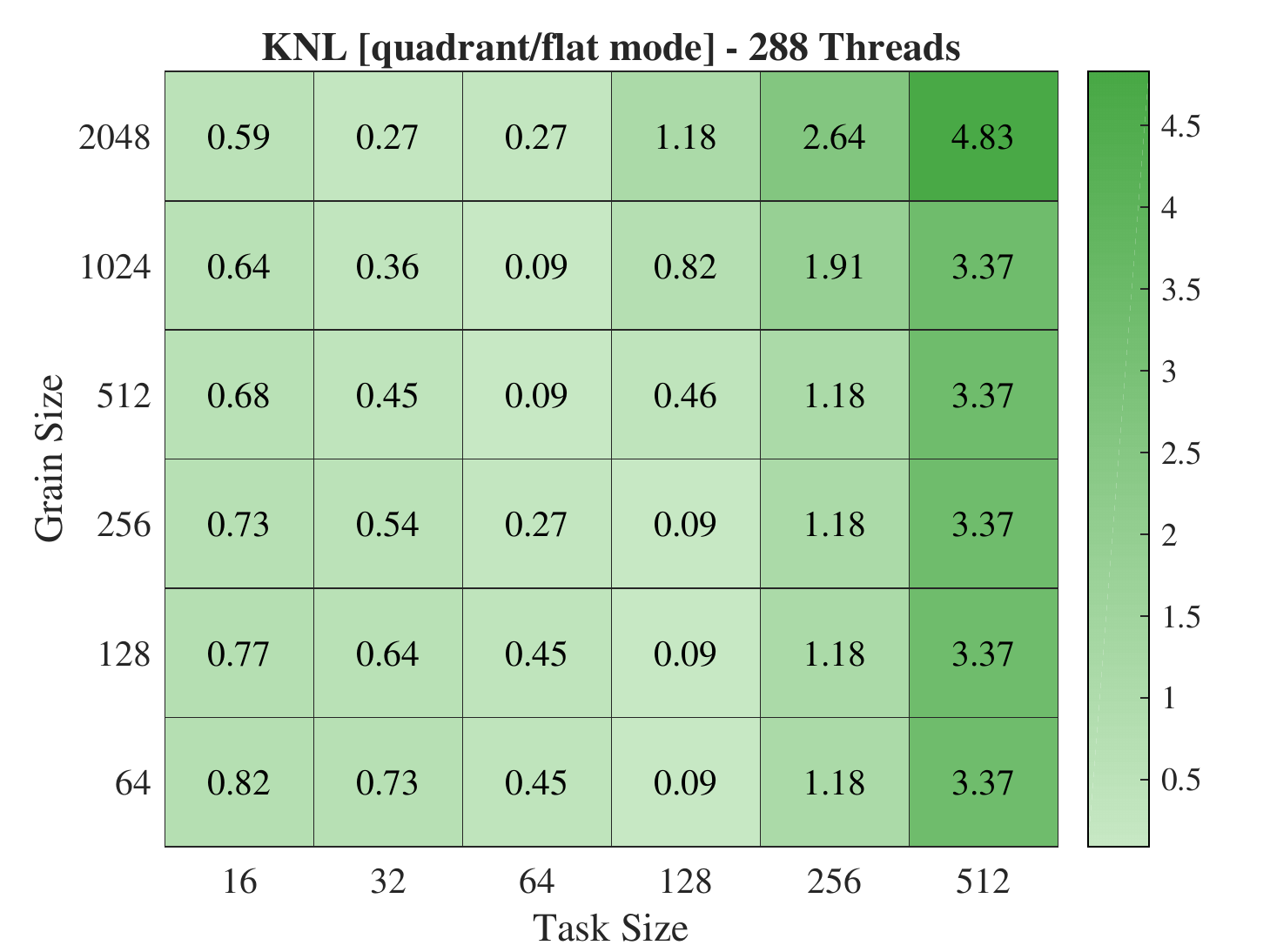}
    \end{center}
  \end{minipage}

  \caption{Tuning the threading parameters performance of the traversal Helmholtz kernel. Performance is normalized by the runtime and the size of the L1/L2/L3 caches on Skylake,
            and the size of the L1/L2 caches on KNL.}
  \label{fig:heat_map1}
\end{figure}

To further illustrate the findings of Figure~\ref{fig:heat_map0}, Figure~\ref{fig:heat_map1} presents the percentage differences between the utilized and available LLC
of a specific architecture (i.e., $36$MB aggregated L2 cache on KNL, and $38$MB L3 exclusive (non-inclusive) cache on Skylake).
We notice a highly accurate prediction of the optimal parameters $(s={64,128,256},c=128)$, which are valid on both architectures.

\subsection{Distributed-memory Parallelism Results}
\label{subsec:perf-distributed-memory}

There are two main goals for this section. The first is to illustrate that the low-level, architectural-specific shared-memory
optimizations employed by this work continue to provide
similar performance benefits as we scale to a large number of compute
nodes.
The second goal, on the other hand, is to demonstrate the performance of the
distributed-memory optimizations that enhance workload
partitioning and communication load balancing.
Also, we depict the scaling properties of our application, establish the scaling limits,
and further study the benefits of data- and thread-level parallelism within a single compute node
in the context of {\bf MPI+Threads+SIMD} hybrid programming paradigm \cite{7161559}.

We observe that for certain heavy MPI
collectives (i.e., \texttt{MPI\_alltoallv} and \texttt{MPI\_Broadcastv}), Cray MPICH
obscurely fails with segmentation faults whenever the amount of data exceeds the $2^{31}$ bytes
limit.
Such failure happens even though the ``count'' argument of MPI routines,
indicates that any array of \texttt{MPI\_Data\_Type} up to $2^{31}$ elements is allowed.
Therefore, in our implementation, we preemptively breakdown the
collectives routines
to multiples of 2 GB of absolute size to eliminate the opportunity for such non-deterministic behaviors. 

In order to wisely exploit the hierarchical interconnect
mode of operation of dragonfly network of Shaheen, we tune the network topology parameters of Slurm.
We trigger hierarchical usage of the network to reduce contention, and we adjust the number
of the network switches for every job request heuristically based on the underlying nature of every experiment \cite{slurm}.
%%%10.1007/10968987_3

\subsubsection{Large-scale Sanity Check}
\label{subsubsec:sanity}

Evaluating the performance of a parallel algorithm can be convoluted. There are many fundamental challenges pertaining to tightly-coupled complicated hardware,
compatibility of the underlying software stacks, theoretical complexity and scaling limits of certain algorithms, and the inadequacy of available
performance counters for different execution stages of certain algorithms. 
In the large-scale context, performance metrics such as speedup and parallel efficiency are natural means of evaluating the performance of certain parallel algorithms
on specific hardware systems. However, these metrics are
essentially related to a predefined base case, and thereafter, they do not reflect the best possible appraisal of the upper- and lower-bound of the algorithmic complexity.

\begin{figure}[h]
    \begin{subfigure}{0.49\textwidth}
      \begin{center}
        \includegraphics[scale=0.43]{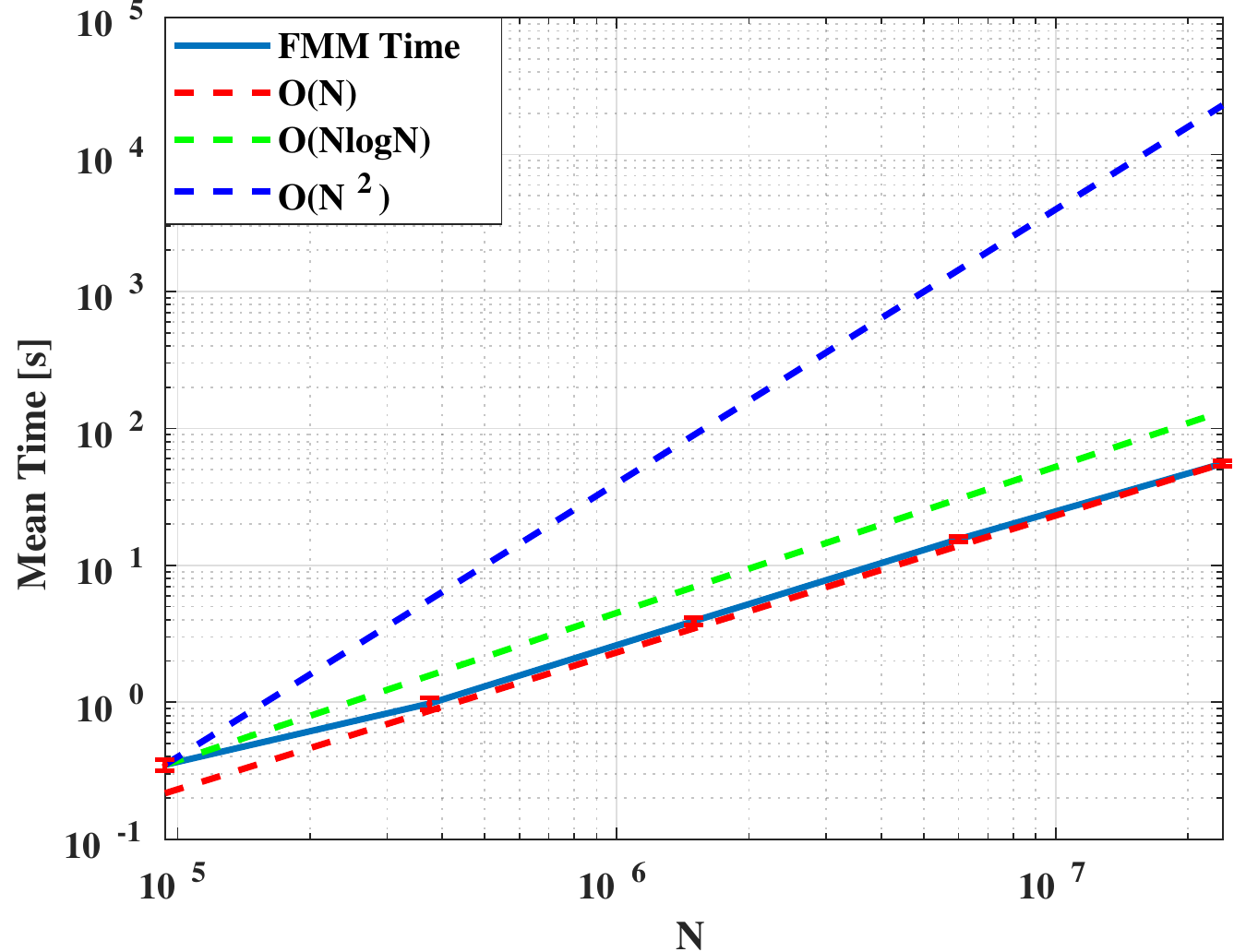}
      \end{center}
    \caption{64 compute nodes of Shaheen}
    \label{fig:datascaling-small}
    \end{subfigure}
    \begin{subfigure}{0.49\textwidth}
      \begin{center}
       \includegraphics[scale=0.43]{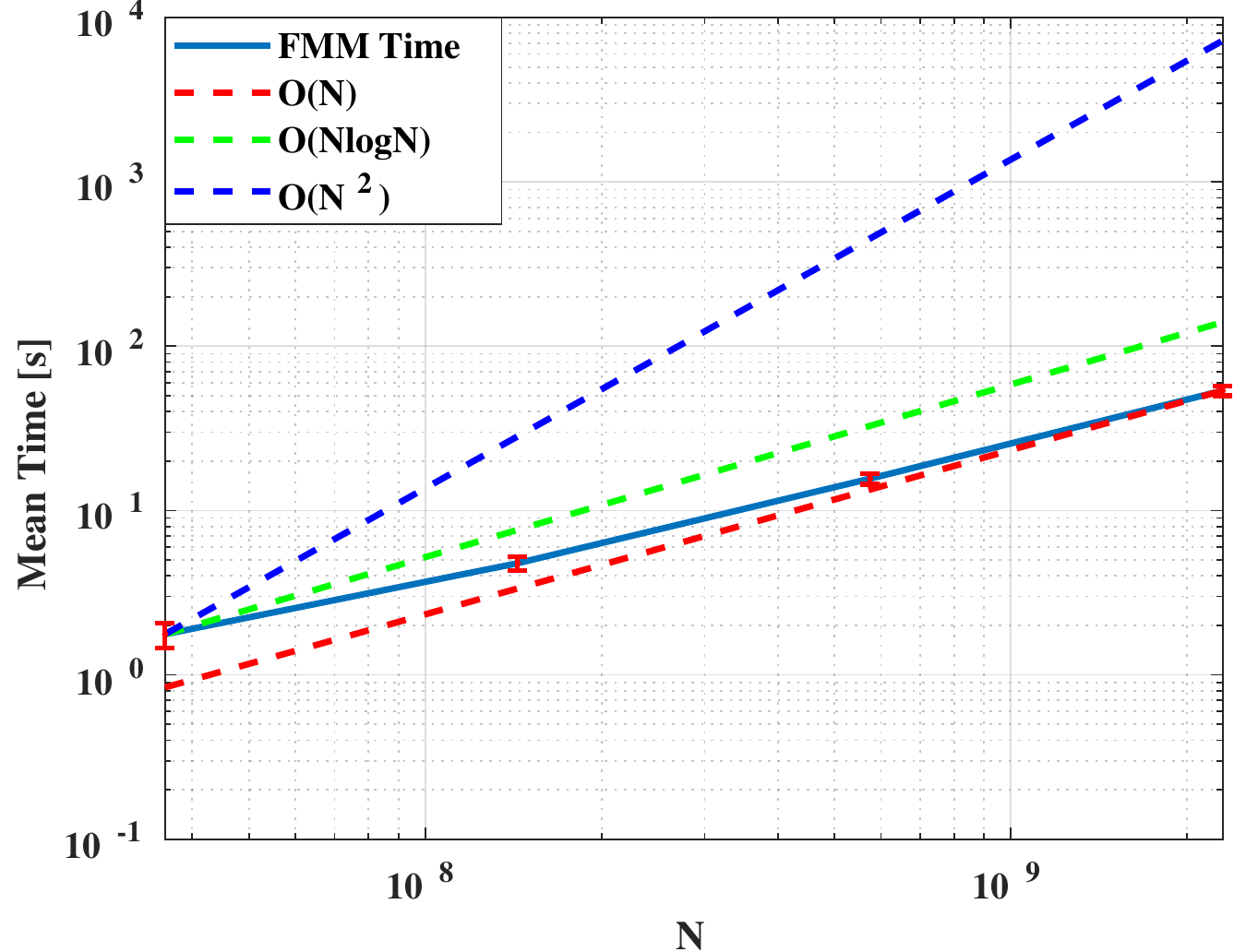}
      \end{center}
    \caption{1,024 compute nodes of Shaheen}
  \label{fig:sdatascaling-large}
    \end{subfigure}\\

  \caption{Scalability results on Shaheen. Performance is normalized by the
          FMM time per linear iteration and the total number of GMRES iterations.}
  \label{fig:datascaling}
\end{figure}

FMM is well-known for reducing the theoretical complexity of matrix-vector product from $O(N^{2})$ to $O(N)$ or ($O(N\log{N})$), depending upon the underlying
traversal algorithm. Therefore, we carry out a ``sanity check'' to verify the theoretical complexity of our FMM Helmholtz implementation at large-scale settings.
Figure~\ref{fig:datascaling} shows the data scaling aspects in practice of FMM kernels with different problem sizes on 64 and 1,024 compute nodes of Shaheen.
The error bars fall within less than 1 standard deviation away from the arithmetic mean of the sample space size.
Furthermore, within a 95\% confidence interval, our FMM Helmholtz implementation achieves a near-linear time complexity. We conclude that for a fixed number of processors, the problem size is linearly proportional to the execution time. Hence, our implementation achieves the theoretical complexity.

\subsubsection{Communication Reduction and Load Balancing Results}
\begin{figure}[h]
     \begin{subfigure}{0.49\textwidth}
       \begin{center}
         \includegraphics[height=4.5cm, width=6.2cm]{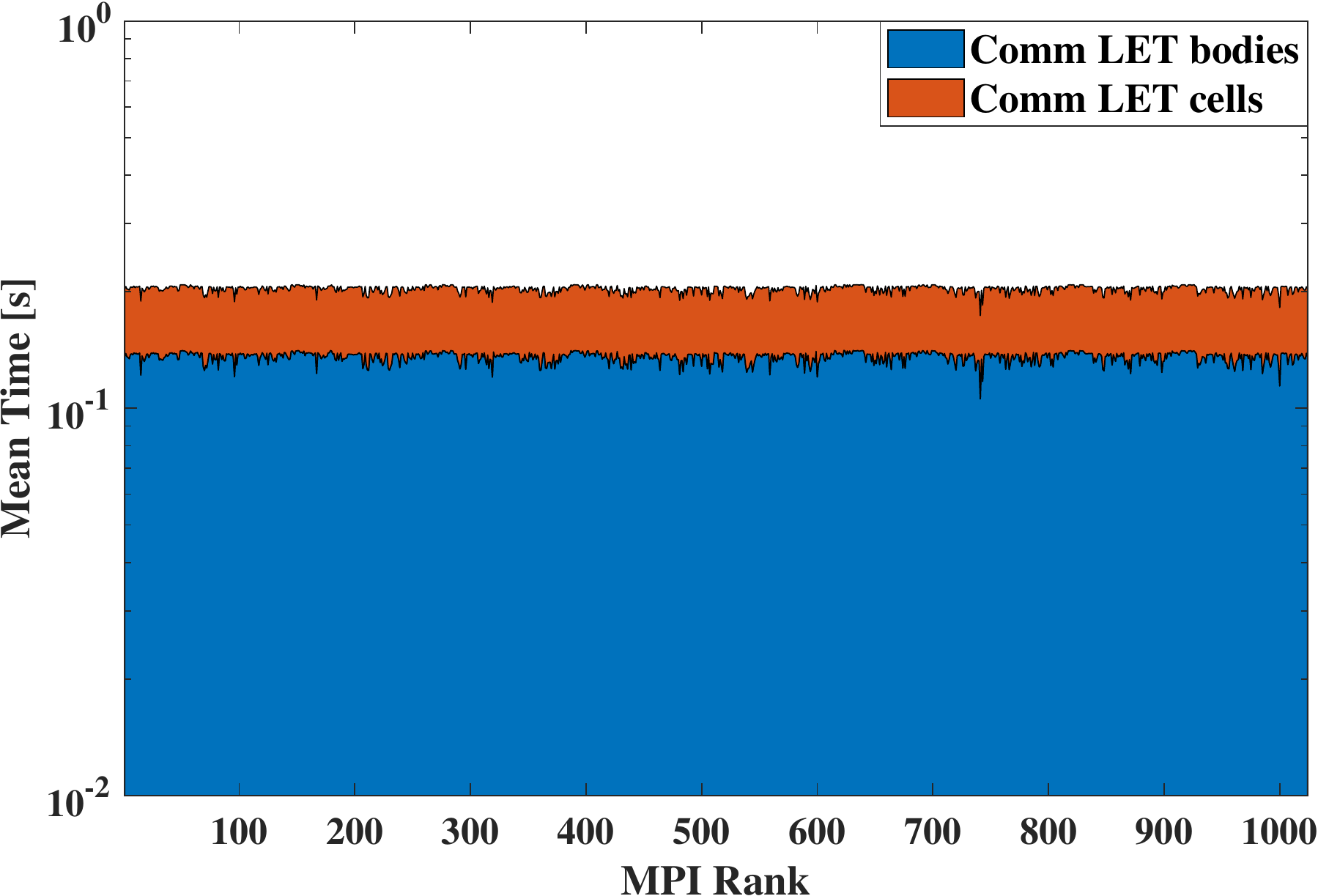}
       \end{center}
     \caption{Initial tree communication time.}
     \label{fig:before_balancing}
     \end{subfigure}
     \begin{subfigure}{0.49\textwidth}
       \begin{center}
        \includegraphics[height=4.5cm, width=6.2cm]{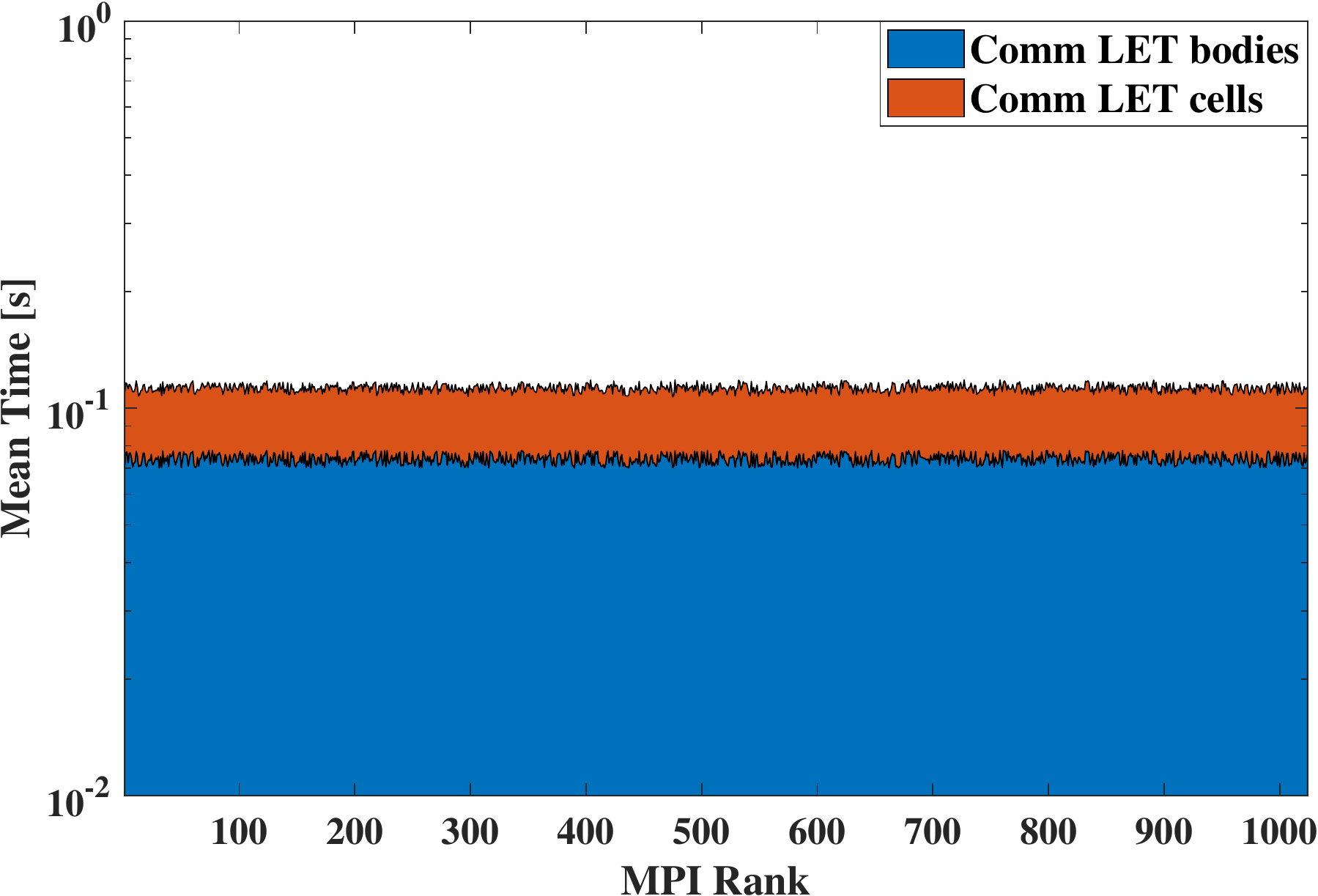}
       \end{center}
     \caption{Time after communication enhancements.}
   \label{fig:after_balancing}
     \end{subfigure}\\
 \begin{center}
   \begin{subfigure}{0.49\textwidth}
 	\begin{center}
 		\includegraphics[height=4.5cm, width=6.2cm]{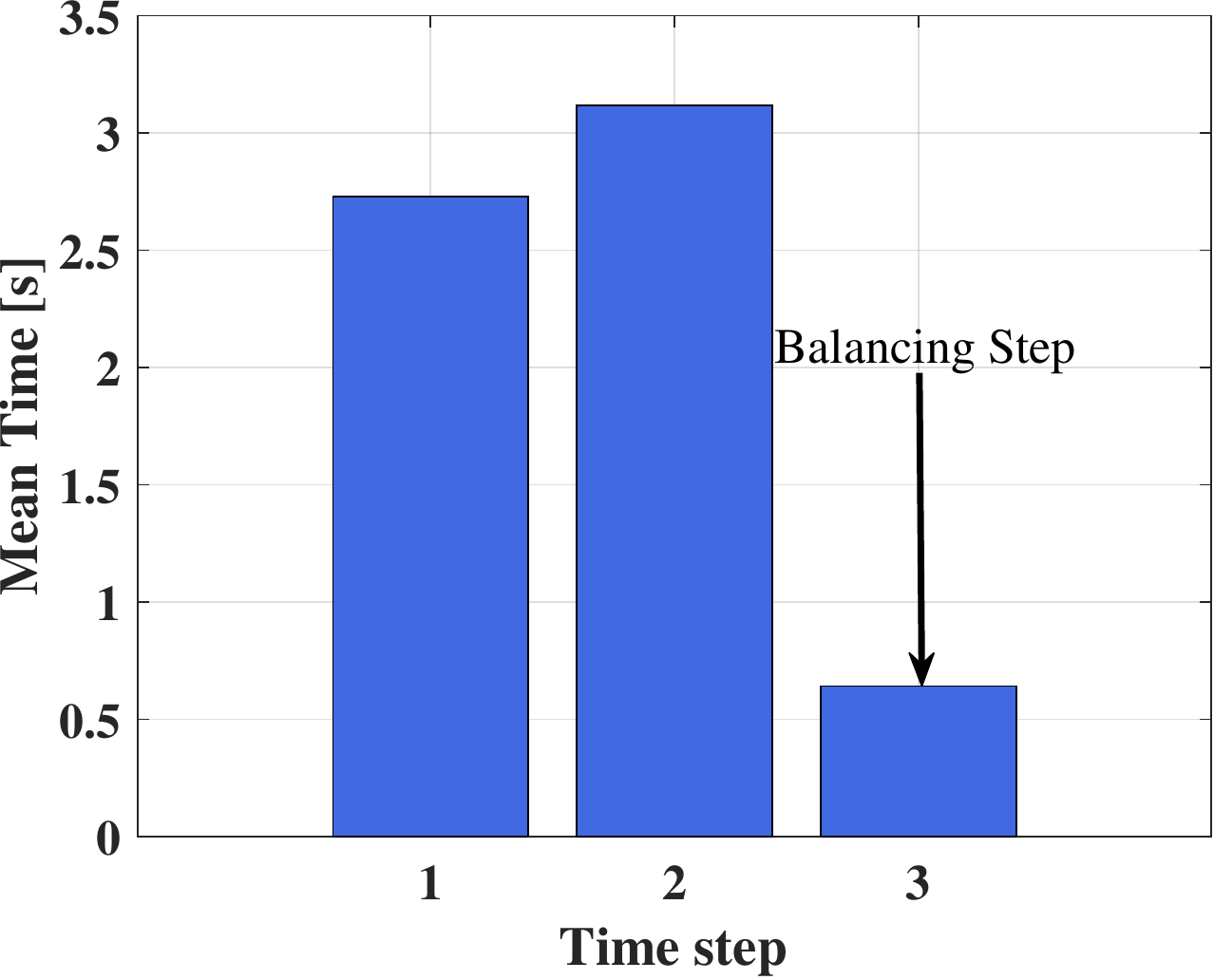}
 	\end{center}
 	\caption{Effect of our optimization techniques on the cumulative communication time on 128 nodes.}
 	\label{fig:cumm_balancing}
 \end{subfigure}
 \end{center}
   \caption{Communication reduction and balancing.}
   \label{fig:load-balance}
 \end{figure}

Communication reducing and balancing using both $\mathcal{HSDX}$ and repartitioning techniques adopted from Section~\ref{subsec:partition_load_balancing} have a vital effect on the cumulative and absolute time of the global tree data exchange. The experiments in Figures~\ref{fig:before_balancing} and~\ref{fig:after_balancing} depict the communication time before and after optimization for $1,024$ nodes and Mesh Q from Table~\ref{tbl:dataset}. The jittery lines in Figure~\ref{fig:before_balancing} represent the imbalance that is smoothed out after repartitioning in Figure~\ref{fig:after_balancing}. We also observe a 1.8x speedup in time (from 0.23 seconds to 0.11 seconds) due to localizing the communication within the Aries Network using $\mathcal{HSDX}$ from Figure~\ref{fig:hsdx}. The cumulative communication time worsens in the case of having smaller node count (e.g., 128 nodes), as we witness a 6-fold improvement when re-balancing is triggered in Figure~\ref{fig:cumm_balancing}.

\label{subsubsec:load_balance}

\subsubsection{Strong and Weak Scalability Studies}
\label{subsubsec:scalability}

In Figure~\ref{fig:scalability}, we present weak and strong scalability results on the Shaheen Cray XC40 supercomputer from 1 compute node
with 32 hardware cores up to 6,144 compute nodes with 196,608 hardware cores on different mesh refinements. The
largest problem involves more than 2 billion DoF.

The cost of our large-scale wave scattering solver is overwhelmingly dominated by the cost of the
underlying FMM Helmholtz kernels that are implicitly invoked by every GMRES iteration
to perform a fast matrix-vector product.

\begin{figure}[h]
    \begin{subfigure}{0.49\textwidth}
      \begin{center}
        \includegraphics[scale=0.42]{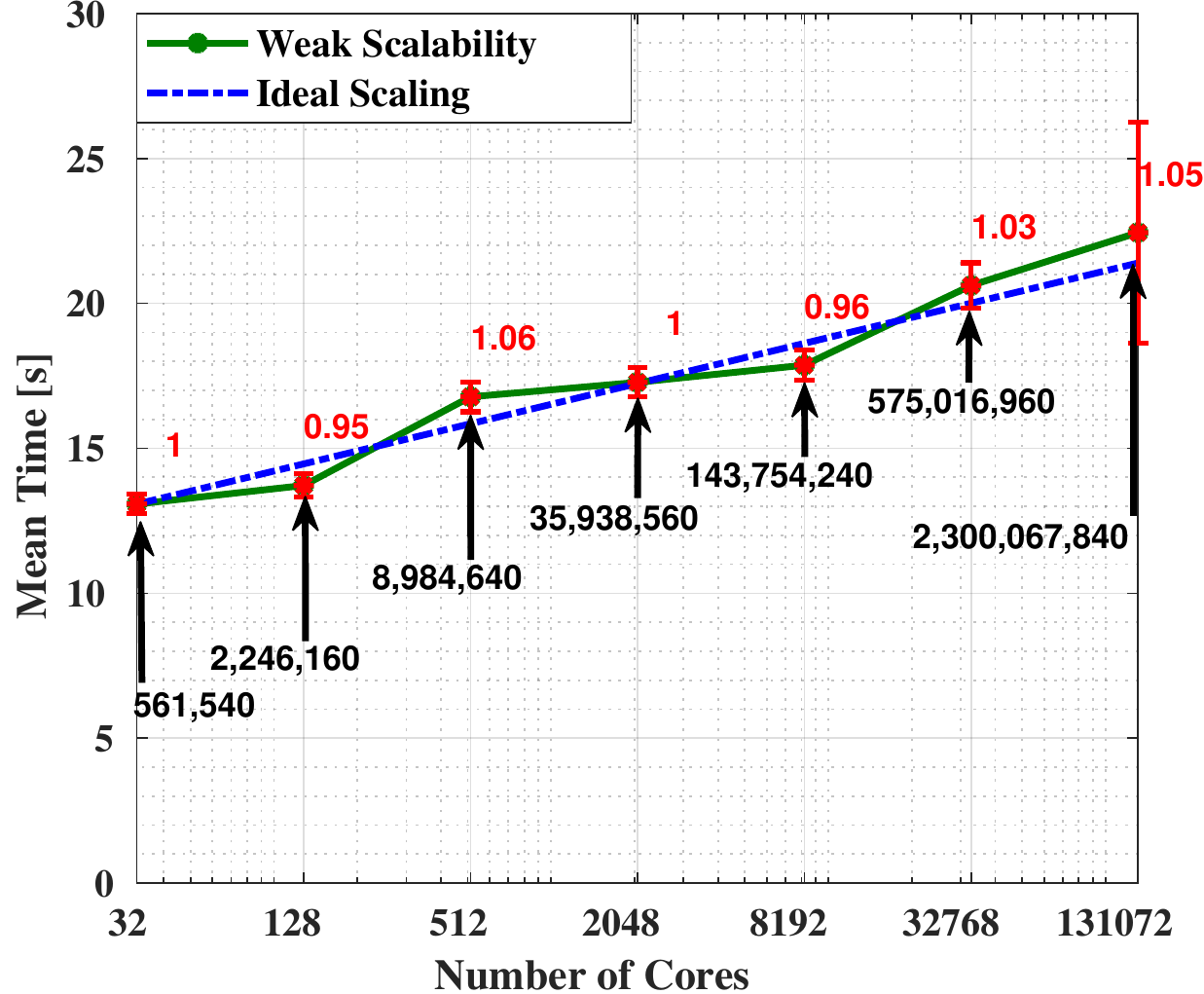}
      \end{center}
    \caption{}
    \label{fig:weak_scalability}
    \end{subfigure}
    \begin{subfigure}{0.49\textwidth}
      \begin{center}
        \includegraphics[scale=0.41]{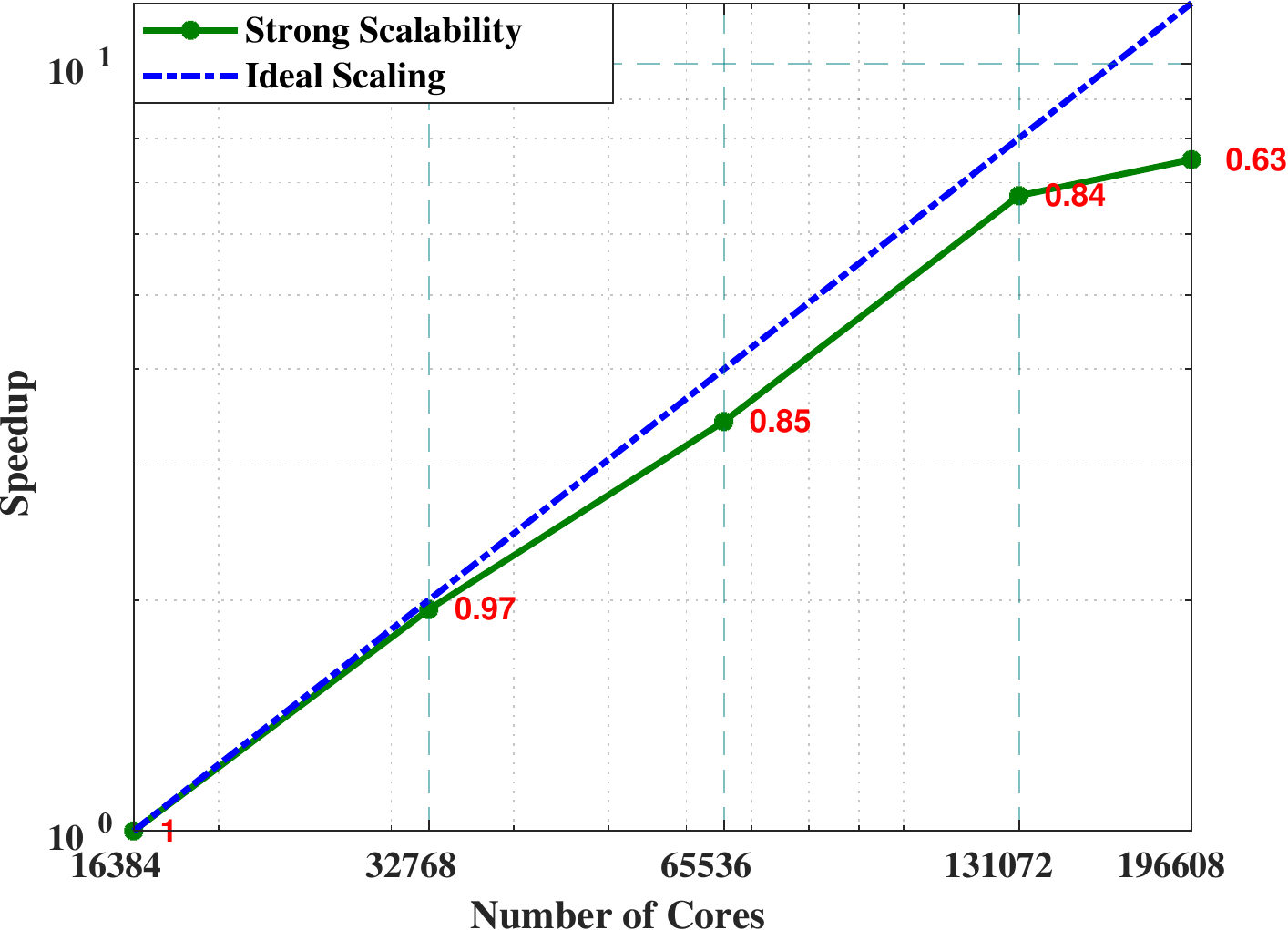}
      \end{center}
    \caption{2,300,067,840 DoF}
  \label{fig:strong_scalability_4096}
    \end{subfigure}\\
    \begin{subfigure}{0.49\textwidth}
      \begin{center}
        \includegraphics[scale=0.41]{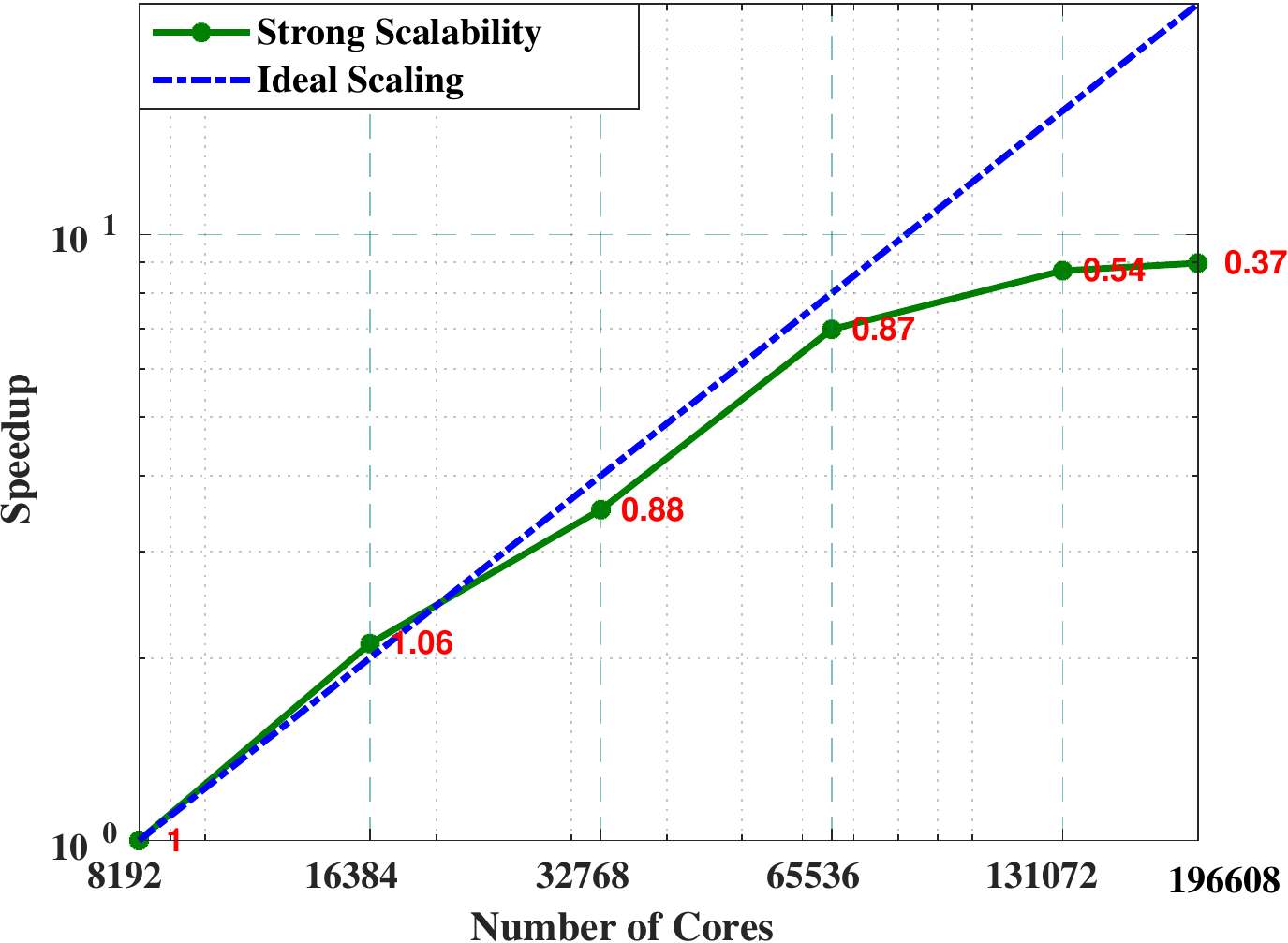}
      \end{center}
      \caption{575,016,960 DoF}
      \label{fig:strong_scalability_1024}
    \end{subfigure}
    \begin{subfigure}{0.49\textwidth}
      \begin{center}
        \includegraphics[scale=0.41]{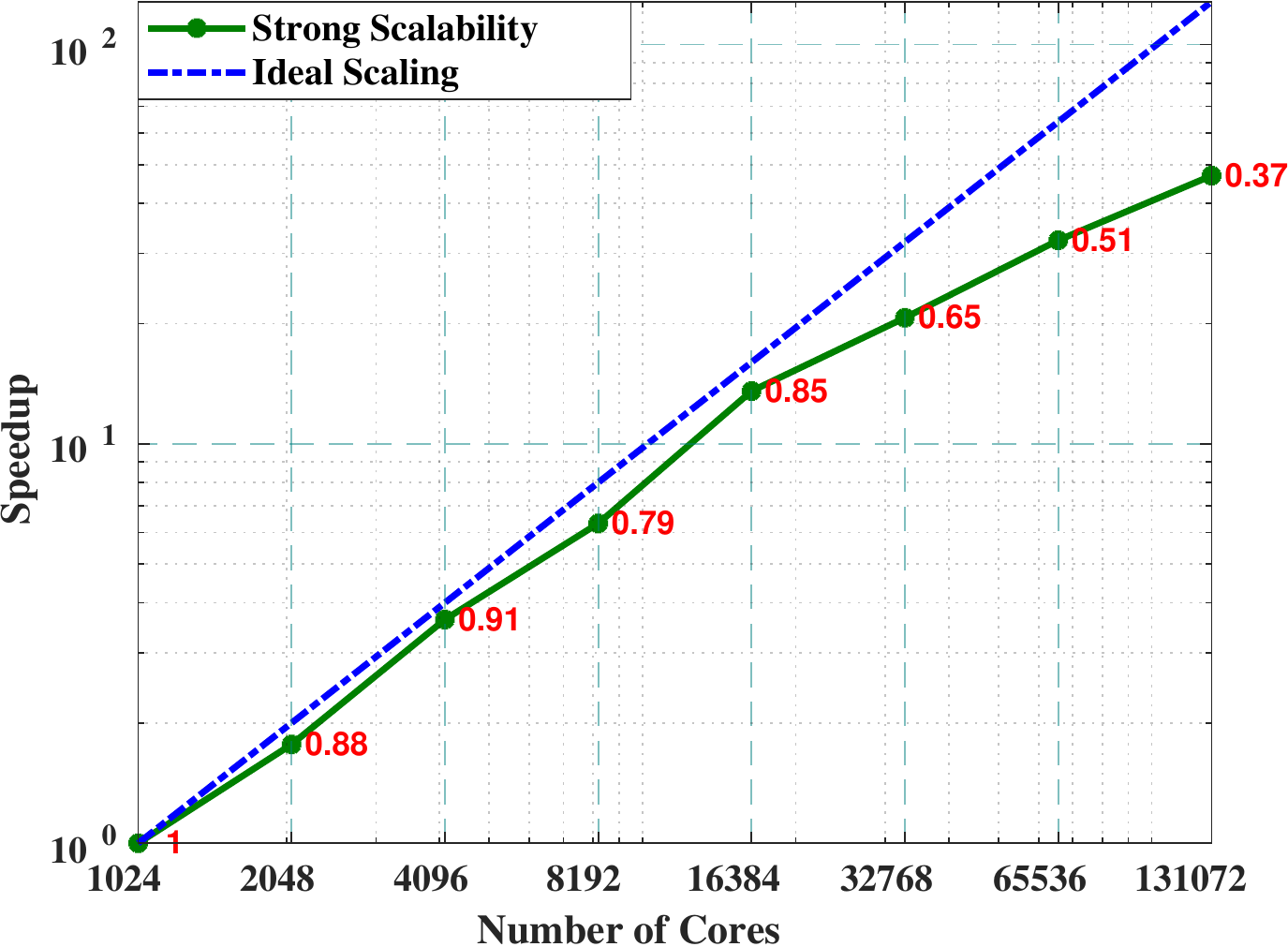}
      \end{center}
       \caption{143,754,240 DoF}
      \label{fig:strong_scalability_256}  
    \end{subfigure}\\
    \begin{subfigure}{0.49\textwidth}
      \begin{center}
        \includegraphics[scale=0.41]{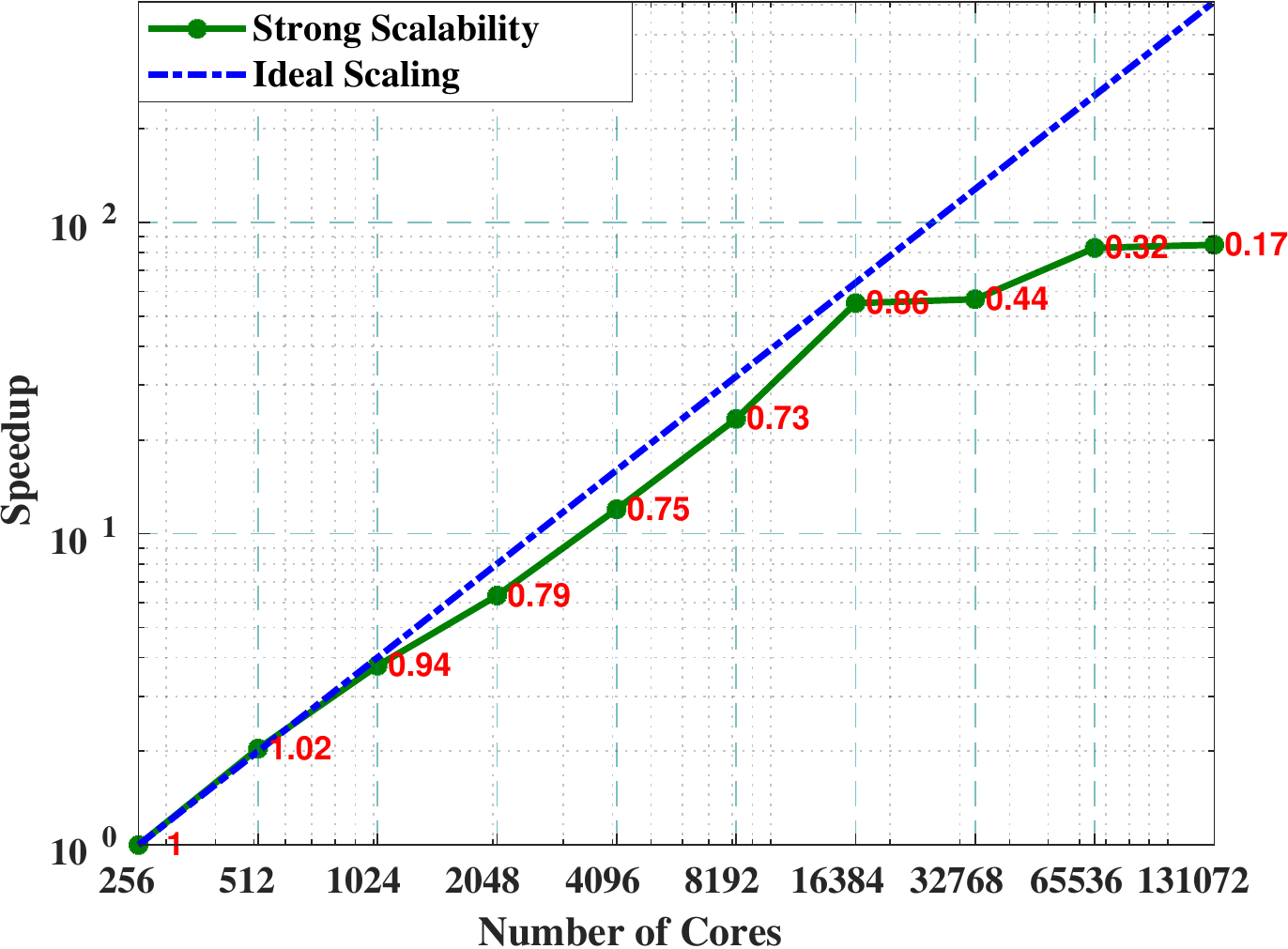}
      \end{center}
      \caption{35,938,560 DoF}
      \label{fig:strong_scalability_64}
    \end{subfigure}
    \begin{subfigure}{0.49\textwidth}
      \begin{center}
        \includegraphics[scale=0.41]{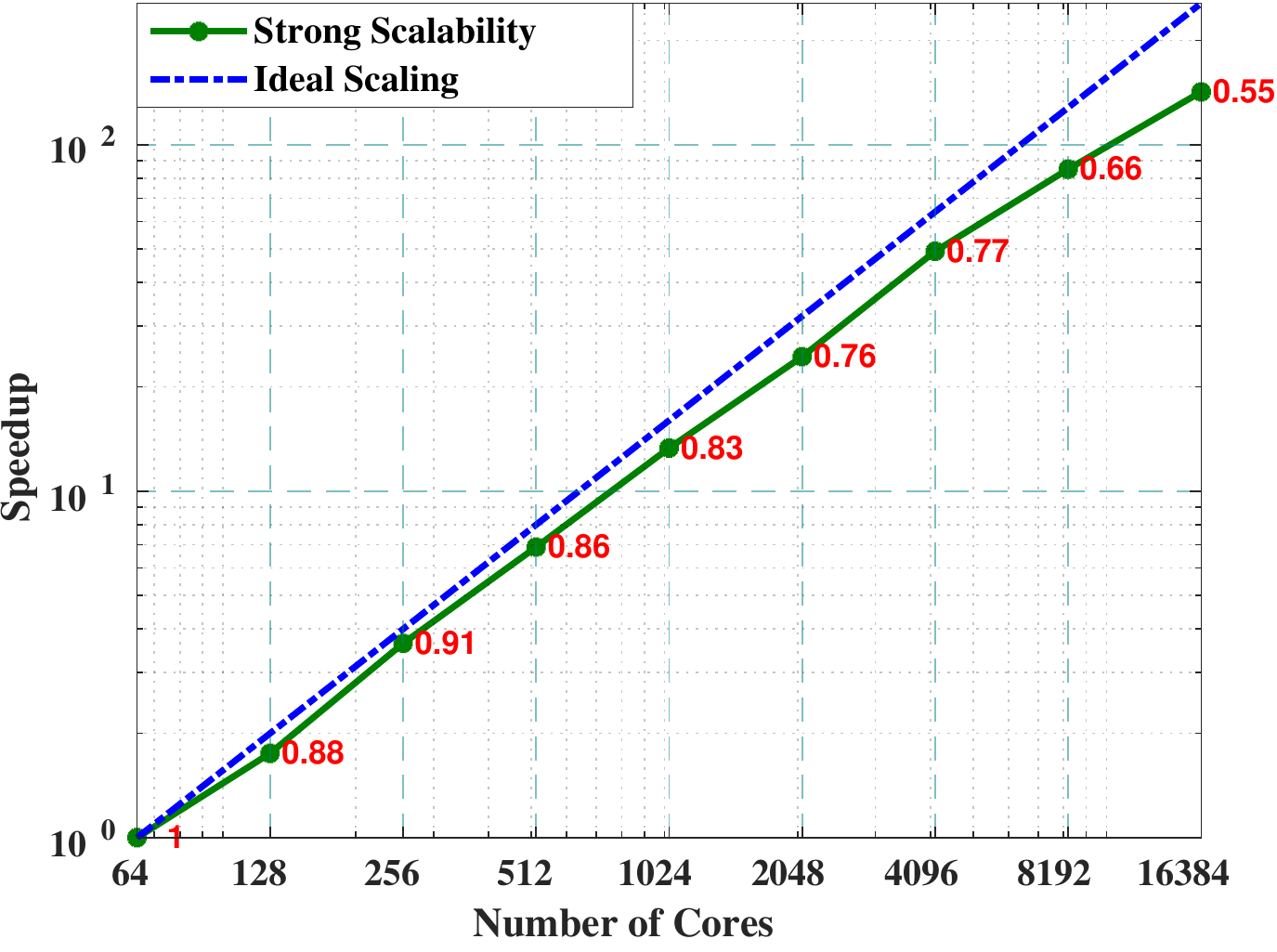}
      \end{center}  
      \caption{8,984,640 DoF}
      \label{fig:strong_scalability_16}
    \end{subfigure}\\
    \begin{subfigure}{0.49\textwidth}
      \begin{center}
        \includegraphics[scale=0.41]{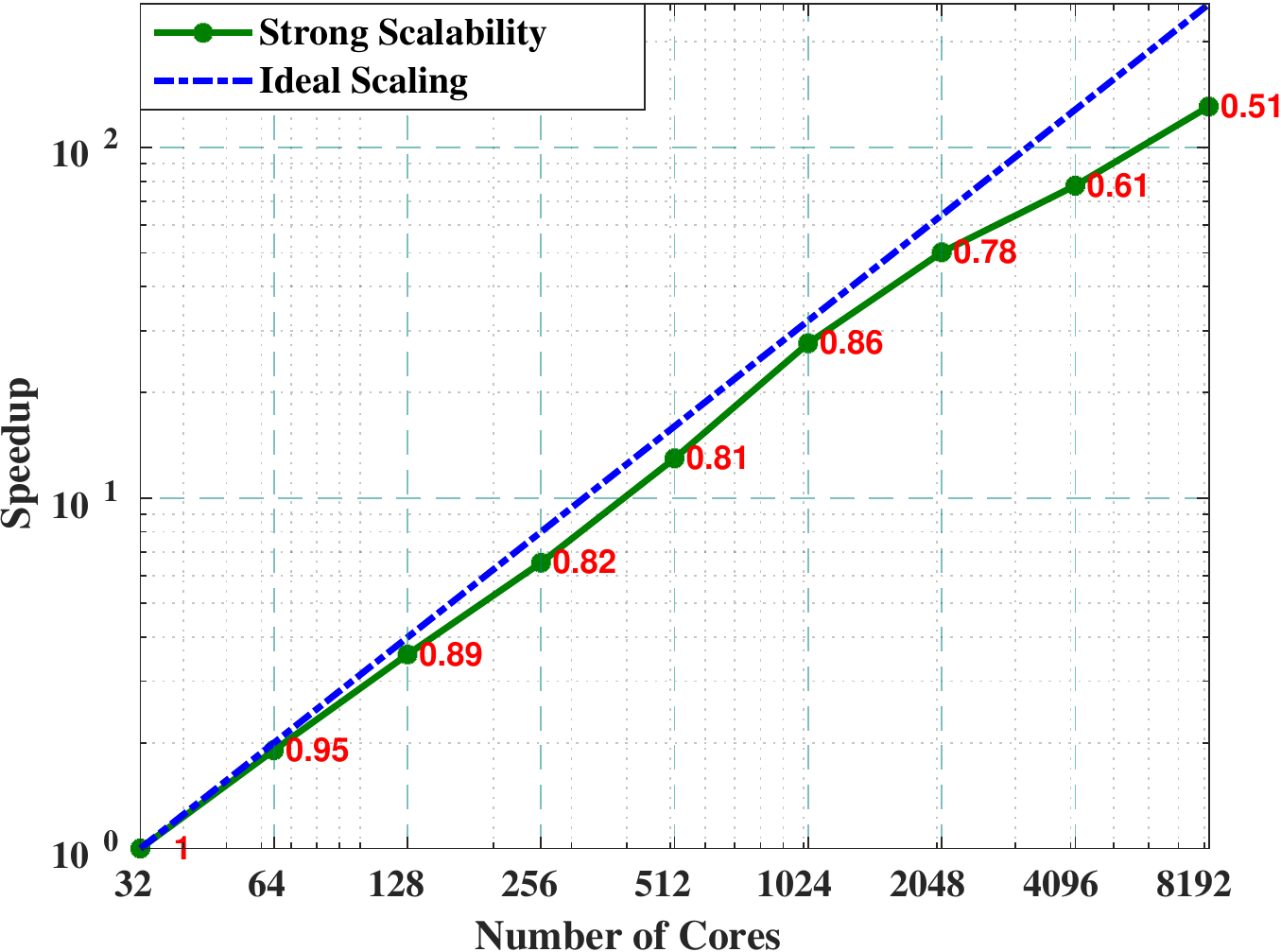}
      \end{center}
       \caption{2,246,160 DoF}
      \label{fig:strong_scalability_4}
    \end{subfigure}
    \begin{subfigure}{0.49\textwidth}
      \begin{center}
        \includegraphics[scale=0.41]{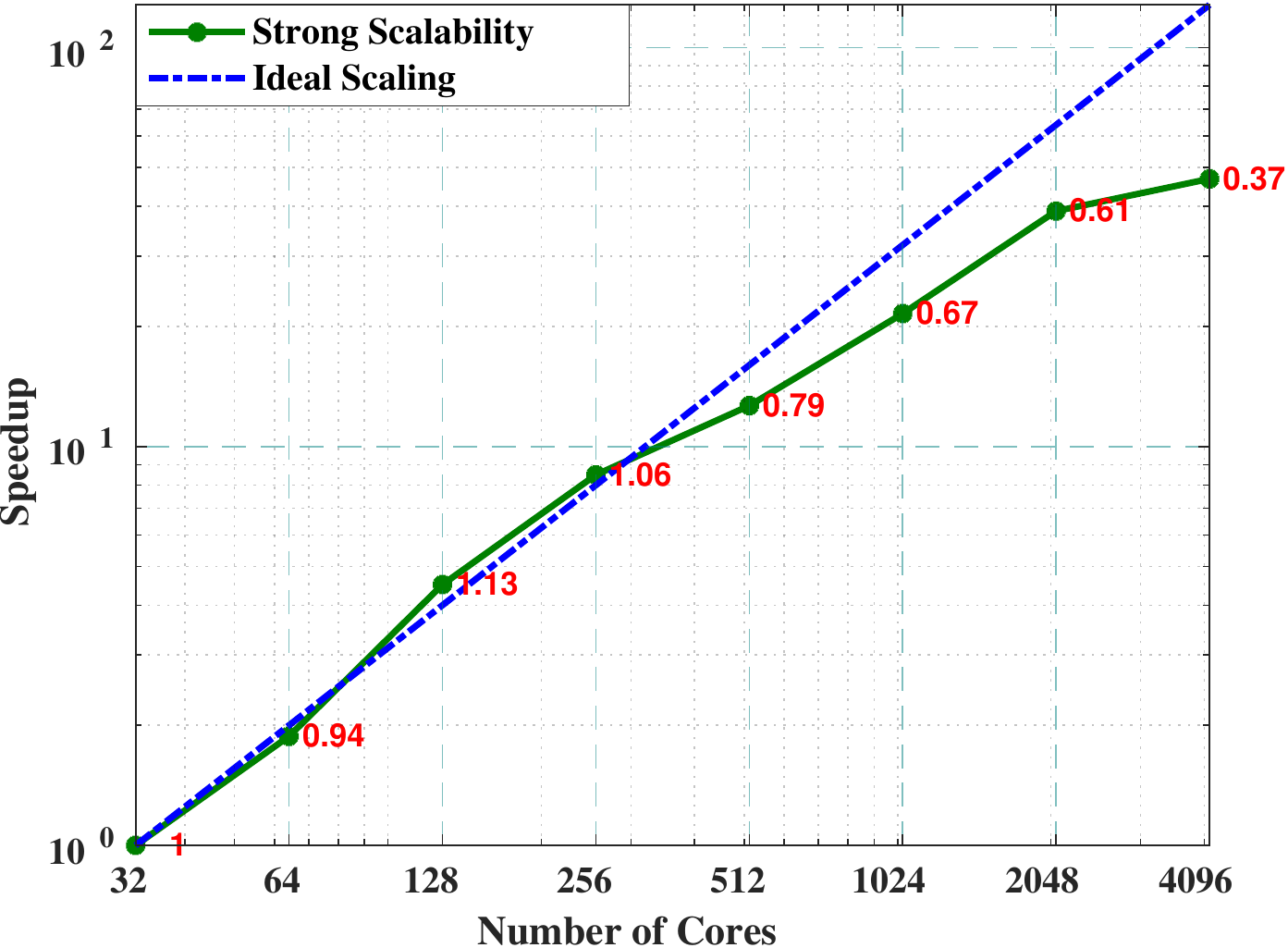}
      \end{center}  
      \caption{561,540 DoF}
      \label{fig:strong_scalability_1}
    \end{subfigure}\\

  \caption{Weak and strong scalability results on Shaheen from 1 to
          6,144 Intel Haswell compute nodes. Performance is normalized by the
          FMM time per linear iteration and the total number of GMRES iterations.
          Numbers along the graph lines (red) indicate efficiency with respect to the
          ideal speedup (efficiency baseline is the smallest compute node count).}
  \label{fig:scalability}
\end{figure}

{\bf Weak Scalability Study:} The main scalability result is the weak scalability (see Figure~\ref{fig:weak_scalability}),
in which we report the FMM time across different mesh refinements.
Table~\ref{tbl:weak_scalability}, shows the weak scalability experimental settings.
At every mesh refinement step, we increase the frequency with the same refinement factor (i.e., a multiple of 4), so that 
we circumvent over-decomposition, and legitimize the necessity of adding extra compute resources in light of the scattering object complexity and wavelengths.

\begin{table}[h]
\centering
  \caption{Weak Scalability results on Shaheen from 1 to 6,144 Intel Haswell compute nodes. Performnce is the FMM
          time-to-solution (residuals converged to 1.0e-4 relative 2-norm residual accuracy).}
  \label{tbl:weak_scalability}
  \begin{tabular}{|c|c|c|c|c|c|}
    \hline
    Cores       & Memory {[}GB{]}   & Frequency {[}KHz{]} & Mesh & Time {[}S{]}   & Iterations \\ \hline
    32          & 128               & 1.5                 & H    & $2.2753e3$     & 175        \\ \hline
    128         & 512               & 6                   & M    & $2.4693e3$     & 180        \\ \hline
    512         & 2,048             & 24                  & N    & $3.1037e3$     & 185        \\ \hline
    2,048       & 8,192             & 96                  & O    & $3.2825e3$     & 190        \\ \hline
    8,192       & 32,768            & 384                 & P    & $3.5737e3$     & 200        \\ \hline
    32,768      & 131,072           & 1,536               & Q    & $4.5982e3$     & 223        \\ \hline
    131,072     & 524,288           & 6,144               & R    & $5.7452e3$     & 256        \\ \hline
  \end{tabular}
\end{table}

Assessing an FMM-based accelerated solver requires a careful consideration of the fact that the most optimal communication scales as
$O(\log{P})$ \cite{Abduljabbar2017_2}. Hence, given the communication complexity, we achieve near-optimal parallel efficiency of FMM per linear iteration
as we refine the mesh, and respectively, increase the hardware resources and the frequency.
In addition to illustrating the effect of the global FMM tree communication on the scalability, Table~\ref{tbl:dragonfly} presents the
critical points, in which we could experience performance degradation based on the Shaheen's network.
The intra-node communication is expected to slow down as we move up across different network units, with hops being the costliest. 
Despite the fact that the communication and the computations are overlapped in our FMM implementation, the communication effects cannot be completely neglected.
Furthermore, the weak scalability results of Figure~\ref{fig:weak_scalability} manifests a logarithmic growth
in the amount of the diagonal and off-diagonal data exchange.
Nonetheless, since we employ our optimized communication protocol, namely $\mathcal{HSDX}$, all tree data exchanges are restricted only by the neighboring compute nodes.
Thus, the overhead of moving data across the dragonfly all-to-all network groups of Shaheen does not affect parallel efficiency, which is
represented by the red numerical labels in the Figure~\ref{fig:scalability}.

\begin{table}[h]
  \centering
  \caption{Characteristics of Shaheen's XC40 Dragonfly Network.}
  \label{tbl:dragonfly}
  \begin{tabular}{|c|c|c|c|c|c|}
    \hline
    Level & Hardware/Network Unit     & Nodes  & Cores    & Cores (Overhead)  & Hops  \\ \hline
    1     & Socket                    &  1     & 16       & 32                & N/A   \\ \hline
    2     & NUMA XC40 Node            &  1     & 32       & 64                & N/A   \\ \hline
    3     & Blade                     &  4     & 128      & 256               & 1     \\ \hline
    4     & Chassis                   &  64    & 2,048    & 4,096             & 1     \\ \hline
    5     & Cabinet                   &  192   & 6,144    & 8,192             & 1     \\ \hline
    6     & Local all-to-all Group    &  384   & 12,288   & 16,384            & 1     \\ \hline
    7     & Global all-to-all Group 1 &  2,304 & 74,728   & 131,072           & 2     \\ \hline
    8     & Global all-to-all Group 2 &  6,174 & 197,568  & N/A               & 3     \\ \hline
  \end{tabular}
\end{table}

In conclusion, we weak scale up to 4,096 compute nodes. Since we refine the mesh in multiples of 4, it is
impractical to weak scale to non-power of two numbers of the compute nodes.
However, we manage to solve a roughly 2 billion DoF system in about 25 seconds per GMRES iteration, which is an average of 100 million DoF per second.
Overall, the solver consistently performs an average of 170 to 250 GMRES iterations until convergence at 1.0e-4 relative 2-norm residual accuracy, and between 40 to 60 minutes time-to-solution. To the best of our knowledge,
this is the fastest FMM-accelerated wave scattering solver for oscillatory kernels (i.e., Helmholtz kernels \cite{MLFMA_fast,KIFMM}).

{\bf Strong Scalability Study:} We perform the study on every mesh of the weak scalability individually
(see Figures~\ref{fig:strong_scalability_4096} through~\ref{fig:strong_scalability_1}). The two largest meshes are scaled up to the full number of the available compute nodes of Shaheen (i.e., 6,144).
We carry the strong scaling far enough to show where the stagnation sets in at large scale,
which is expected as the problem size per node gets smaller and the communication time becomes dominate.
Additionally, since the data has to travel across the Dragonfly network units or all-to-all hops (see Table~\ref{tbl:dragonfly}),
the downturn in the parallel efficiency is foreseen.
In other word, as we travel past the first all-to-all group, a consistent performance instability is experienced, especially after
level 6 (i.e., 16,384 cores) in Figures~\ref{fig:strong_scalability_4096},~\ref{fig:strong_scalability_1024},~\ref{fig:strong_scalability_256}, and~\ref{fig:strong_scalability_64}.
Similarly, propagation within a local all-to-all group, up to three cabinets,
may adversely affect the performance. For example, observe the performance beyond 256 hardware cores in Figure~\ref{fig:strong_scalability_1}).
Nonetheless, the performance subtleties in such cases are almost negligible, since the communication is mostly
hidden within the local computations at the lower compute core counts.

\subsection{Convergence Aspects and Numerical Error}
\label{subsec:analytical-and-numerical-solution}

\begin{figure}[h]
  \begin{subfigure}{0.49\textwidth}
    \begin{center}
      \includegraphics[scale=0.45]{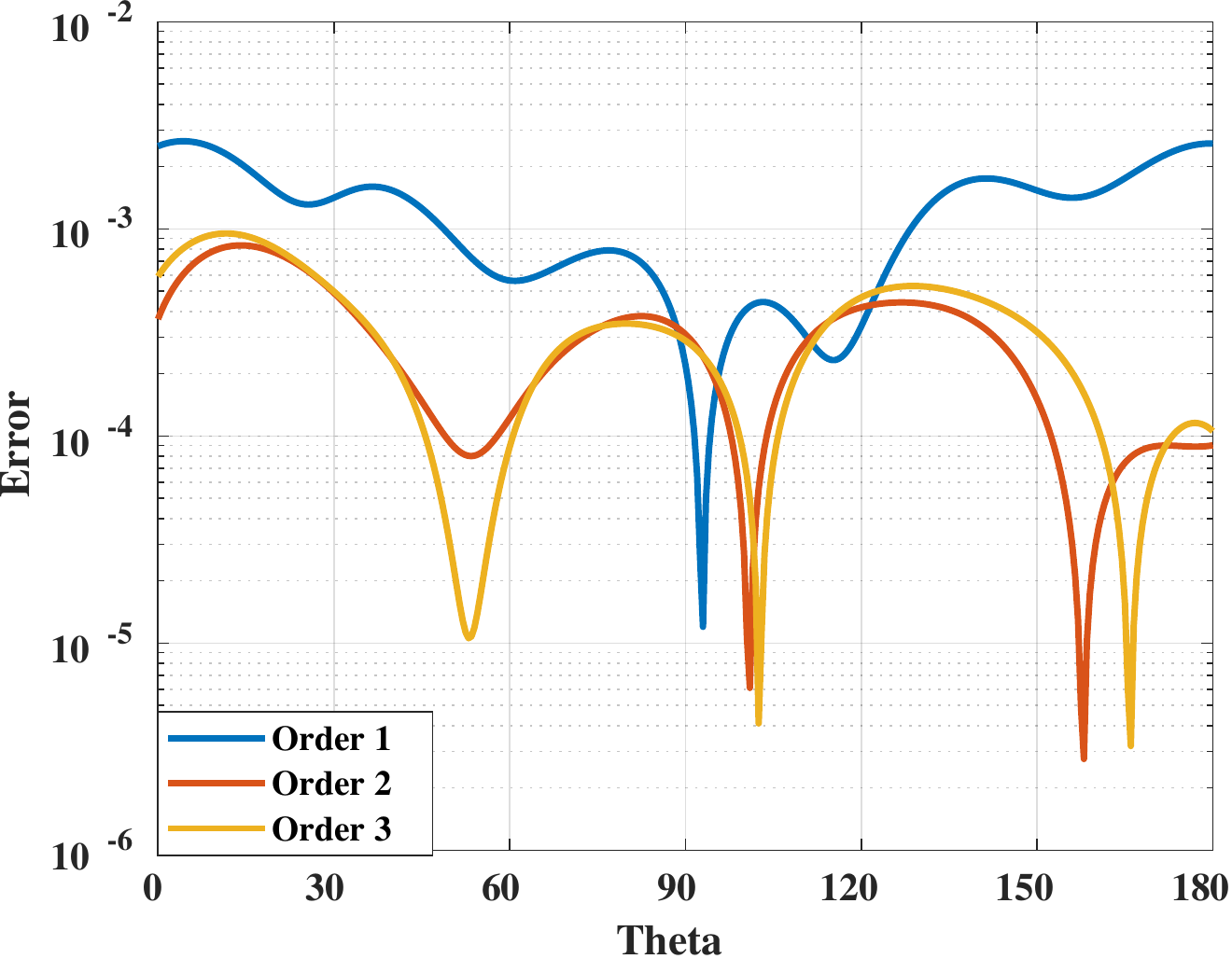}
    \end{center}
    \caption{}
    \label{fig:near_field_vs_mie}
  \end{subfigure}
  \begin{subfigure}{0.49\textwidth}
    \begin{center}
      \includegraphics[scale=0.45]{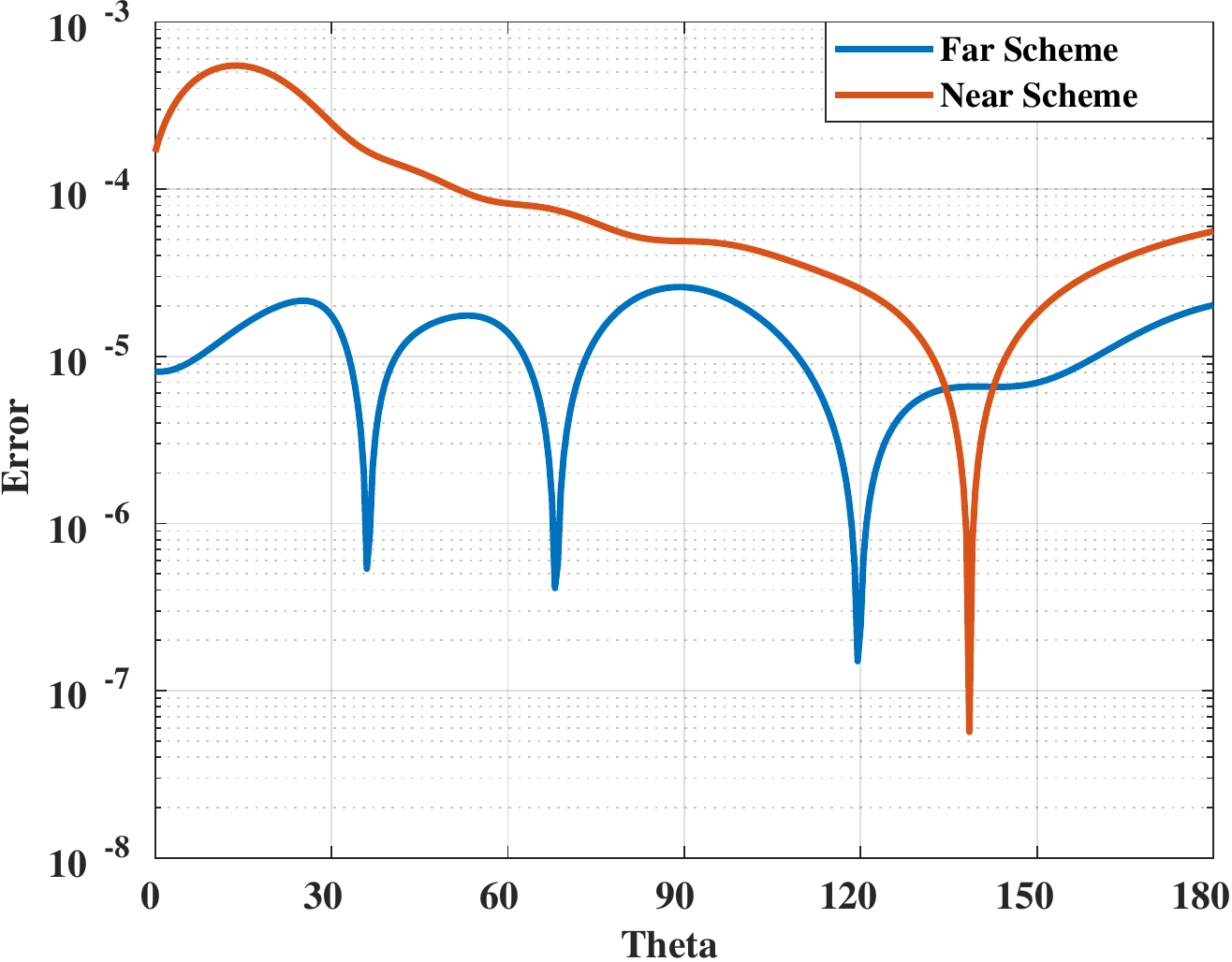}
    \end{center}
    \caption{}
    \label{fig:err_near_field_vs_far}
  \end{subfigure}
  
  \caption{Numerical solution accuracy for an acoustics field arising from 43,644 DoF.}
  \label{fig:numerical}
\end{figure}

This experiment demonstrates the accuracy and efficiency of the developed FMM kernels inside the GMRES iterative solver for acoustic wave scattering.
It is performed for the single layer kernels associated with Helmholtz equation.
Surfaces are discretized with high-order curvilinear-isoparametric quadrilateral elements.
The use of high-order curvilinear elements is crucial for the accurate implementation of a high-order solution.
In addition, applying a high-order scheme to large elements with high-order basis functions is characteristically more accurate,
in which the slope of the error curve becomes steeper as we increase the basis order.
Nevertheless, accuracy may suffer due to the geometry approximation if surfaces are not accurately modeled.
We demonstrate this phenomenon of interest as follows. In Figure~\ref{fig:near_field_vs_mie}, we initially consider the near-field acoustic wave
scattering by a soft sphere of radius $a=1m$. The observation points are $4m$ away from the sphere center, at which
the scattering angle ($\theta$) ranges from: $0^{\circ}-180^{\circ}$. ($\phi$) is artificially to $0^{\circ}$.
As mentioned earlier, the sphere is discretized with $6$ quadrilateral curvilinear elements. ``order-1'', ``order-2'', and ``order-3'' in the legend denote the basis function order of the first-degree three points,
second-degree six points, and third-degree 12 points, respectively.
The higher-order contribution is clearly seen with such $p$-refinement,
where excellent agreement is observed with the analytical solution, and the error is inversely proportional to the basis order.
Figure\ref{fig:err_near_field_vs_far} shows the absolute error for the far and near scattered fields.

% \begin{figure}[h]
%   \centering
%   \includegraphics[scale=0.75]{convergence_64x30m.pdf}
%   \caption{Convergence effects of implementing self and near singularity treatments inside FMM Helmholtz kernel.}
%   \label{fig:convergence}
% \end{figure}

\begin{figure}[h]
\centering
\begin{minipage}{.49\textwidth}
  \centering
  \includegraphics[scale=0.775]{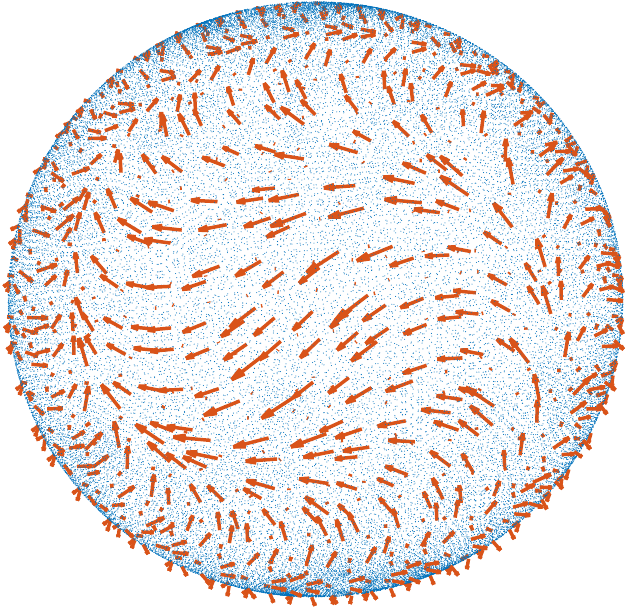}
  \caption{Incident solution velocity field on the spherical surface.}
  \label{fig:solution}
\end{minipage}
\hspace{0.5mm}
\begin{minipage}{.49\textwidth}
  \centering
  \includegraphics[scale=0.43]{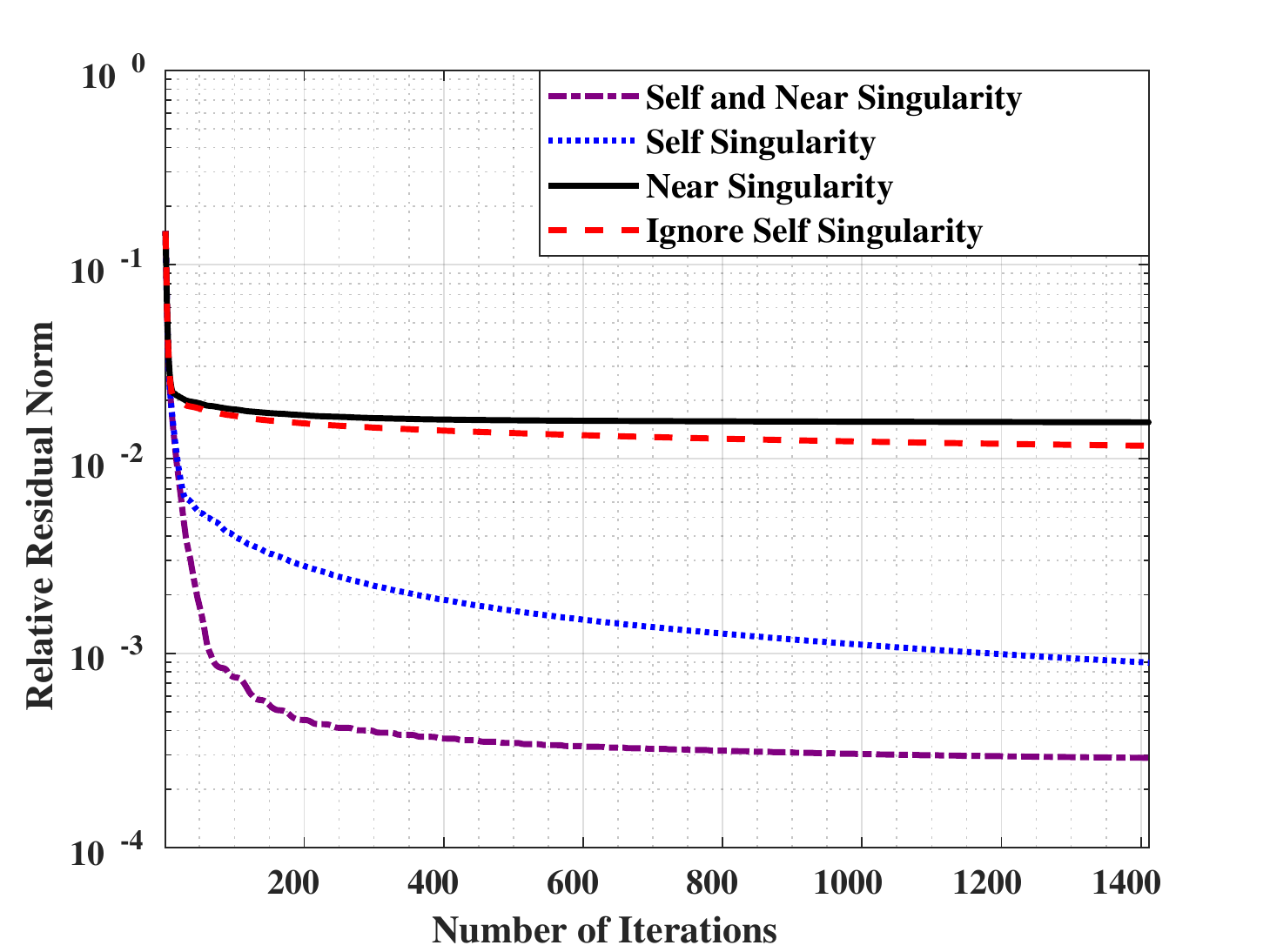}
  %\caption{Convergence effects of implementing self and near singularity treatments inside FMM Helmholtz kernel.}
  \caption{Convergence effects of self and near singularity treatments.}
  \label{fig:convergence}
\end{minipage}
\end{figure}

Figure~\ref{fig:convergence} shows the convergence behavior of the solver where we calculate the far scattered fields interactions by a sphere of radius $1m$ using different singularity treatment modes. The observation points have the same properties as in Figure~\ref{fig:numerical}. The Y-axis shows the number of iterations required for GMRES convergence, whereas the X-axis represents the GMRES relative residual norm. The self-singularity treatment creates higher order Gauss quadrature points around the sources and targets falling exactly on the diagonal, or having [$R<\epsilon$] from geometrical perspective. Such points are ignored in typical FMM implementations. Near-singularity scheme treats points that fall within [$R < near$] radius ($near$ is a code parameter). It can be clearly depicted in Figure~\ref{fig:convergence} that when singularity is ignored, we exhibit a lower convergence rate that is driven and dominated by the singularity of the Green's function. However, when treating only the true singularity, a remarkably improved convergence rate is observed. Finally, when considering the self and near singularity treatment schemes, the relative 2-norm residual accuracy reached 1.0e-4 within just less than 50 iterations. This accuracy corresponds to 1.0e-3 to 1.0e-4
error with respect to the analytical solution, and it cannot be further improved even with less relative
tolerance of the underlying iterative solution. Hence, our GMRES solver is configured to exit at 1.0e-4
relative 2-norm residual accuracy.
In addition, Figure~\ref{fig:solution} simulates the velocity solution vector (Equation~\ref{eq8}) on a spherical plane.

% \begin{figure}[h]
%   \centering
%   \includegraphics[scale=1]{rj_solution.pdf}
%   \caption{Incident solution velocity field on the spherical surface.}
%   \label{fig:solution}
% \end{figure}

%\clearpage

%%%%%%%%%%%%%%%%%%%%%%%%%%%%%%%%%%%%%%%%%%%%%%%%%%%%%%%%%%%%%%%%%%%%%%%%%%%%%%%%%%%%%
%%%%%%%%%%%%%%%%%%%%%%%%%%%%%%%%%%%%%%%%%%%%%%%%%%%%%%%%%%%%%%%%%%%%%%%%%%%%%%%%%%%%%

%%%%%%%%%%%%% \section{Related Work}
%%%%%%%%%%%%% \label{sec:related_work}

%%%%%%%%%%%%%%%%%%%%%%%%%%%%%%%%%%%%%%%%%%%%%%%%%%%%%%%%%%%%%%%%%%%%%%%%%%%%%%%%%%%%%
%%%%%%%%%%%%%%%%%%%%%%%%%%%%%%%%%%%%%%%%%%%%%%%%%%%%%%%%%%%%%%%%%%%%%%%%%%%%%%%%%%%%%

\section{Concluding Remarks and Future Work}
\label{sec:conclusion}

We summarize the progress of this contribution towards migrating
frequency-domain scattering to contemporary extreme architectures
on a path to exascale:

\begin{itemize}[noitemsep, leftmargin=*]
  \item
  Implementation of a scattering solver for complex 3D Helmholtz including:
  \begin{itemize}
    \item
    A numerical iterative linear solver based upon GMRES that uses FMM as a fast and accurate matrix-vector multiplication accelerator.
    \item
    A nontrivial singularity treatment for near and self integration points.
  \end{itemize}
  \item
  Low-level optimization and fine-tuning for the shared-memory performance that address
  different emerging HPC architectures, including:
  \begin{itemize}
    \item
    Optimal data-level parallelism through efficient vectorization.
    \item
    Specific data structure allocation and striding to enhance the compiler optimization. 
    \item
    Yardsticks for writing SIMD-friendly, low-level FMM-Helmholtz kernel codes.
    \item
    Fine-tuned thread-level parallelism of the traversal kernels of FMM.
    \item
    Predictive performance models for selecting the task-based parameters
    to tune the threading performance.
  \end{itemize}
  \item
  Efficient large-scale architecture-specific and algorithm-aware implementation for distributed-memory parallelisms that manifests:
  \begin{itemize}
    \item
    Adaptive nonuniform partitioning and load balancing schemes.
    \item
    Scalable communication reducing protocols.
    \item
    Large-scale performance model based on Cray's dragonfly network topology.
  \end{itemize}
  \item
  Strong and weak scalability studies up to 6,144 compute nodes of
  a Cray XC40 with 196,608 Intel Haswell cores.
  \item
  Data scalability study to demonstrate the theoretical complexity of FMM on distributed memory systems.
  \item
  Solution of a 2 billion DoF systems of high-order curvilinear triangular patches of a spherical mesh in about 60 minutes time-to-solution, and roughly 250 GMRES linear iterations
  to achieve a relative 2-norm residual accuracy of 1.0e-4.
  \item
  Near-optimal parallel efficiency on Shaheen for both weak and strong scalability studies.
  \item
  Single precision floating point performance of Skylake and KNL
  of more than 60\% of theoretical peak,
  which represents a speedup of 5.4x over the out-of-the-box compilation.
\end{itemize}

For future considerations, we are extending this work to address the
performance challenges of more heterogeneous HPC architectures (e.g., GPU).
In addition, we are planning to study different network topologies of various supercomputer architectures, including the bleeding-edge Intel Omni-Path network architecture.
Our solver implementation will be further extended to include different highly nonuniform domains.
We are also adapting to more complex geometries, for instance, wing-fuselage configuration emulated by two intersecting ellipsoids.

%%%%%%%%%%%%%%%%%%%%%%%%%%%%%%%%%%%%%%%%%%%%%%%%%%%%%%%%%%%%%%%%%%%%%%%%%%%%%%%%%%%%%
%%%%%%%%%%%%%%%%%%%%%%%%%%%%%%%%%%%%%%%%%%%%%%%%%%%%%%%%%%%%%%%%%%%%%%%%%%%%%%%%%%%%%

\section*{Acknowledgments}
Support in the form of computing resources was provided by
KAUST Extreme Computing Research Center,
KAUST Supercomputing Laboratory,
KAUST Information Technology Research Division,
Intel Parallel Computing Centers,
and Cray Supercomputing Center of Excellence.
In particular, the authors are very appreciative to Bilel Hadri of KAUST Supercomputer
Laboratory for his great help and support throughout scalability experiments
on the Shaheen supercomputer.

%%%%%%%%%%%%%%%%%%%%%%%%%%%%%%%%%%%%%%%%%%%%%%%%%%%%%%%%%%%%%%%%%%%%%%%%%%%%%%%%%%%%%
%%%%%%%%%%%%%%%%%%%%%%%%%%%%%%%%%%%%%%%%%%%%%%%%%%%%%%%%%%%%%%%%%%%%%%%%%%%%%%%%%%%%%

\bibliographystyle{siamplain}
\bibliography{main.bib}
\end{document}